\documentclass{article}

\usepackage{arxiv}

\usepackage[utf8]{inputenc} 
\usepackage[T1]{fontenc}    
\usepackage{hyperref}       
\usepackage{url}            
\usepackage{booktabs}       
\usepackage{amsfonts}       
\usepackage{nicefrac}       
\usepackage{microtype}      
\usepackage{lipsum}

\usepackage{authblk}

\usepackage{cite}
\usepackage[pdftex]{graphicx}
	 \usepackage{epstopdf}
   \DeclareGraphicsExtensions{.pdf,.jpg,.jpeg,.png,.eps}
\usepackage[abs]{overpic}
\usepackage{amsfonts}
\usepackage{tikz}
\usetikzlibrary{arrows,shapes,positioning,fit,patterns,decorations.pathmorphing}
\usepackage[export]{adjustbox}
\usepackage{array}
\usepackage[caption=false,font=normalsize,labelfont=sf,textfont=sf,subrefformat=parens,labelformat=parens]{subfig}
\usepackage{etex}

\usepackage{threeparttable}
\usepackage{booktabs}
\usepackage{multirow}
\usepackage{tabularx}

\usepackage{tablefootnote}

\usepackage{amsmath,amssymb}
\usepackage{color}
\usepackage{xspace}
\usepackage[normalem]{ulem}
\usepackage{bm}

\makeatletter
\DeclareRobustCommand\onedot{\futurelet\@let@token\@onedot}
\def\@onedot{\ifx\@let@token.\else.\null\fi\xspace}
\def\etal{~et~al\onedot}
\def\eg{e.g\onedot} 
\def\ie{i.e\onedot}

\makeatother

\def\clap#1{\hbox to 0pt{\hss #1\hss}}%
\def\initials#1{\protect\clap{\smash{\raisebox{1.4ex}{\tiny{\textsf{\textit{#1}}}}}}}%
\makeatletter
\newcommand{\EDIT}[4][]{\protect\@ifundefined{hidecomments}{%
  \strut{\color{#3}{\hspace{0pt}\initials{#2}\protect\sout{#1}{~#4}}}%
  }{#4}}
\newcommand{\NOTEboxed}[3]{\protect\@ifundefined{hidecomments}{%
  {\begin{center}\fbox{\parbox{0.97\linewidth}{\protect\EDIT{#1}{#2}{#3}}}\end{center}}
  }{}}
\newcommand{\COMM}[3]{\protect\@ifundefined{hidecomments}{%
  {\protect\EDIT{#1}{#2}{#3}}
  }{}}
\newcommand{\DefAuthor}[2] 
{%
  \expandafter\newcommand\csname #1edit\endcsname[2][]{\protect\EDIT[##1]{#1}{#2}{##2}}
  \expandafter\newcommand\csname #1\endcsname[1]{\protect\COMM{#1}{#2}{[##1]}}
  \expandafter\newcommand\csname #1boxed\endcsname[1]{\NOTEboxed{#1}{#2}{##1}}
}
\definecolor{dfltgreen}       {rgb}{0.0,0.5,0.0}
\definecolor{dfltred}         {rgb}{0.7,0.0,0.0}
\newcommand{\REVadd}[1]{\protect\@ifundefined{hidecomments}{%
  \strut{\color{dfltgreen}{#1}}}{#1}}
\newcommand{\REVedit}[2][]{\protect\@ifundefined{hidecomments}{%
  \strut{\color{dfltred}{\protect\sout{#1}}\color{dfltgreen}{~#2}}}%
  {#2}}
\makeatother

\def \to {,\ldots ,}

\DeclareMathOperator*{\sinc}{sinc}

\newcommand{\ReferNoIE}{\textbf{Refer}}
\newcommand{\ReconstNoIE}{\textbf{Reconst}}
\newcommand{\ReferXu}{\textbf{Refer\_Xu}}
\newcommand{\ReconstXu}{\textbf{Reconst\_Xu}}
\newcommand{\ReferLRG}{\textbf{Refer\_LR-G}}
\newcommand{\ReconstLRG}{\textbf{Reconst\_LR-G}}
\newcommand{\ReferLRXu}{\textbf{Refer\_LR-Xu}}
\newcommand{\ReconstLRXu}{\textbf{Reconst\_LR-Xu}}

\title{Computational Image Enhancement for Frequency Modulated Continuous Wave (FMCW) THz Image}

\author[1, 2]{Tak~Ming~Wong}
\author[1, 3]{Matthias~Kahl}
\author[1, 3]{Peter~Haring~Bol\'ivar}
\author[1, 2]{Andreas~Kolb}

\affil[1]{Center for Sensor Systems (\emph{ZESS}), University of Siegen, 57076 Siegen, Germany}
\affil[2]{Computer Graphics and Multimedia Systems Group, University of Siegen, 57076 Siegen, Germany}
\affil[3]{Institute for High Frequency and Quantum Electronics (\emph{HQE}), University of Siegen, 57068 Siegen, Germany}

\date{January 23, 2019}

\begin{document}
\maketitle


\begin{abstract}

In this paper, a novel method to
enhance Frequency Modulated Continuous Wave (FMCW) THz imaging
resolution beyond its diffraction limit is proposed. Our method comprises two stages. Firstly, we
reconstruct the signal in depth-direction using a $\sinc$-envelope,
yielding a significant improvement in depth estimation and signal parameter extraction. 
The resulting high precision depth
estimate is used to deduce an accurate reflection intensity THz image.
This image is fed in the second stage of our method to a 2D blind deconvolution
procedure, adopted to enhance the lateral THz image resolution beyond the diffraction limit. 
Experimental data acquired with a FMCW system operating at 577 GHz with a bandwidth of 126 GHz shows that the proposed method enhances the lateral resolution by a factor of
2.29 to 346.2um with respect to the diffraction limit. 
The depth accuracy is 91um. Interestingly, the lateral resolution enhancement achieved with this blind deconvolution concept leads to better results in comparison to conventional gaussian deconvolution. Experimental data on a PCB resolution target is presented, in order to quantify the resolution enhancement and to compare the performance with established image enhancement approaches. The presented technique allows exposure of the interwoven fibre reinforced embedded structures of the PCB test sample. 

\end{abstract}



\section{Introduction}
\label{s:intro}

Since its original inception in the early nineties~\cite{hu1995imaging}, 
THz imaging has demonstrated a very large potential for contact-free analysis, 
nondestructive testing and stand-off detection in a wide variety of application fields, 
such as semiconductor industry, biology, medicine, material analysis, quality control, and security~\cite{siegel2002terahertz,chan2007imaging,Jansen2010}. 
In many of these application fields, THz imaging is competing with established imaging methodologies, 
such as optical inspection or X-ray imaging.
Compared with imaging in the optical or X-ray parts of the electromagnetic spectrum, 
THz imaging is significantly limited in its spatial resolution due to the substantially longer
wavelength of the associated frequencies.

A wide range of technological approaches to realize THz imaging systems have been demonstrated, 
including femtosecond laser based scanning systems~\cite{hu1995imaging,cooper2011thz}, 
synthetic aperture systems~\cite{mcclatchey2001time,ding2013thz}, and hybrid systems~\cite{kahl2012}. 
Several of these approaches using pulses or using frequency modulated THz signals, 
allow to sense temporal or phase shifts to the object's surface, making 3D THz imaging possible.

Despite the fact that in most of these approaches THz imaging is performed close to the diffraction limit, 
the comparatively low spatial resolution associated with the THz radiation wavelengths significantly hampers the application range. 
There is a huge interest to increase the spatial resolution of this approach beyond the diffraction limit, 
in order to make this technique competitive in comparison to established methods such as X-ray imaging~\cite{Ahi15}.

THz imaging below the diffraction limit is an emerging area~\cite{chan2007imaging}, 
which can roughly be classified into two alternatives: by system enhancements or by computational approaches. 
System enhancements include for example, interferometric sensing~\cite{johnson2001interferometric} to increase the depth sensitivity in THz time-of-flight imaging, 
or near-field sensing approaches~\cite{chen2003terahertz} which demonstrate a nanometer scale lateral resolution by circumventing the diffraction limit. 
Computational image enhancement techniques aim at improving the resolution by utilizing numerical procedures and additional signal or system information, 
\eg from overlapping oversampled THz imaging signals, without introducing additional equipment effort and cost.

Depending on the THz image acquisition mode, there are several alternative approaches for computational image enhancement. 
THz imaging superresolution (also referred to as high-resolution or image restoration) is often associated to spatial resolution enhancement in the xy-direction~\cite{xu2014high,li2008super,ding2010high,hou2014enhancing,ahi2016developing}. 
In contrast, depth resolution enhancement is associated to improvement in azimuth direction (z-direction)~\cite{walker2012terahertz,chen2010frequency,takayanagi2009high}.\par

In this paper, we propose a novel method to enhance THz image resolution beyond diffraction limit, attaining a lateral (xy) resolution increase and a depth (z) accuracy increase. 
The concept is demonstrated for a Frequency Modulated Continuous Wave (FMCW) THz scanning system operating at $514-640$GHz, 
but can certainly be adapted to other THz imaging techniques. 
Our approach comprises the following technical contributions:
\begin{itemize}
\item Parameter extraction by complex signal fitting model in z-direction for each pixel which allows for the acquisition of \emph{non-planar} targets, incorporating
  \begin{itemize}
  \item an accurate estimation of the per-pixel distance to the object surface, and
  \item a proper reconstruction of the reflection intensity as a measure for the object's material-properties.
  \end{itemize}
\item Based on the accurately reconstructed reflection intensity, we are able to apply more complex, 
state-of-art 2D blind deconvolution techniques in order to improve the spatial xy-resolution beyond what is achievable with traditional (\eg gaussian kernel) deconvolution procedures.
\end{itemize}

Sections \ref{s:system} and \ref{s:method} give a brief overview of
the THz 3D imaging system and the proposed method, respectively.  The
details of the curve fitting procedure are described in
Sec.~\ref{s:parameter_extraction}.  In Sec.~\ref{s:evaluation}, the
evaluation of the computational result of the proposed method are depicted.\par


\section{Prior Work}
\label{s:prior}

The majority of prior research focuses on the lateral resolution of 2D THz images, where the Lucy-Richardson deconvolution algorithm~\cite{lucy1974iterative,richardson1972bayesian} is one of the most frequently used methods. 
Knobloch\etal~\cite{knobloch2002medical} firstly applied Lucy-Richardson deconvolution on THz images.
Xu\etal~\cite{xu2014high} proposed a THz time-domain spectroscopy (THz-TDS) image high-resolution reconstruction model 
incorporating a 2D wavelet decomposition and a Lucy-Richardson deconvolution to reconstruct a high-resolution THz image from four low-resolution images 
and to reconstruct a high-resolution image from single degraded 2D low-resolution image. Li\etal~\cite{li2008super} proposed to use the Lucy-Richardson deconvolution algorithm 
for a coherent THz 2D imaging system. Ding\etal~\cite{ding2010high} used the Lucy-Richardson deconvolution for a THz reflective 2D imaging system.\par

In addition to a 2D deconvolution algorithm, Hou\etal~\cite{hou2014enhancing} proposed a method to enhance THz image quality by applying a Finite Impulse Response Filter in time domain. 
Ahi and Anwar~\cite{ahi2016developing} proposed a deconvolution method based on their THz image generation model for a THz far-field 2D imaging system. Xie\etal~\cite{xie2013adaptive} proposed to use a Markov Random Field (MRF) model for THz superresolution.\par

For the THz blur kernel estimation, Ding\etal~\cite{ding2013thz} proposed to use a Range Migration Algorithm (RMA). Gu\etal~\cite{gu2013three} proposed to use a physical model for THz point spread function (PSF) as a Gaussian kernel based on the optical parameters of a THz-TDS system.\par

To enhance the depth accuracy, Walker\etal~\cite{walker2012terahertz} proposed a time domain deconvolution method to increase the time domain zero padding factor for a THz reflection imaging. 
Chen\etal~\cite{chen2010frequency} proposed a hybrid Frequency-Wavelet Domain Deconvolution (FWDD) method to improve calculation of impulse response functions. 
Takayanagi\etal~\cite{takayanagi2009high} proposed a deconvolution method based on Wiener filtering in the time domain. 
Dong\etal~\cite{dong2017terahertz} proposed a sparse deconvolution method of impulse response function on multi-layered structure.\par

In the context of THz 3D imaging reconstruction method Shift Migration (PSM) algorithm to reconstruct a 3D image, Sun\etal~\cite{sun2014fast} proposed to extend PSM as Enhanced Phase Shift Migration (EPSM) to improve computational efficiency. 
Liu\etal~\cite{liu2015fast} proposed a 3D Wavenumber Scaling Algorithm (3D WSA) to reconstruct the entire 3D holographic image.\par

To our best knowledge, the method in this paper is the first attempt that combines
depth accuracy with lateral resolution enhancement in order to achieve
a jointly improved spatial resolution and accuracy in both, $xy$- and $z$-direction.


\section{Overview of Terahertz 3D imaging system}
\label{s:system}

\begin{figure}[!t]
	\centering
	\includegraphics[width=0.9\textwidth]{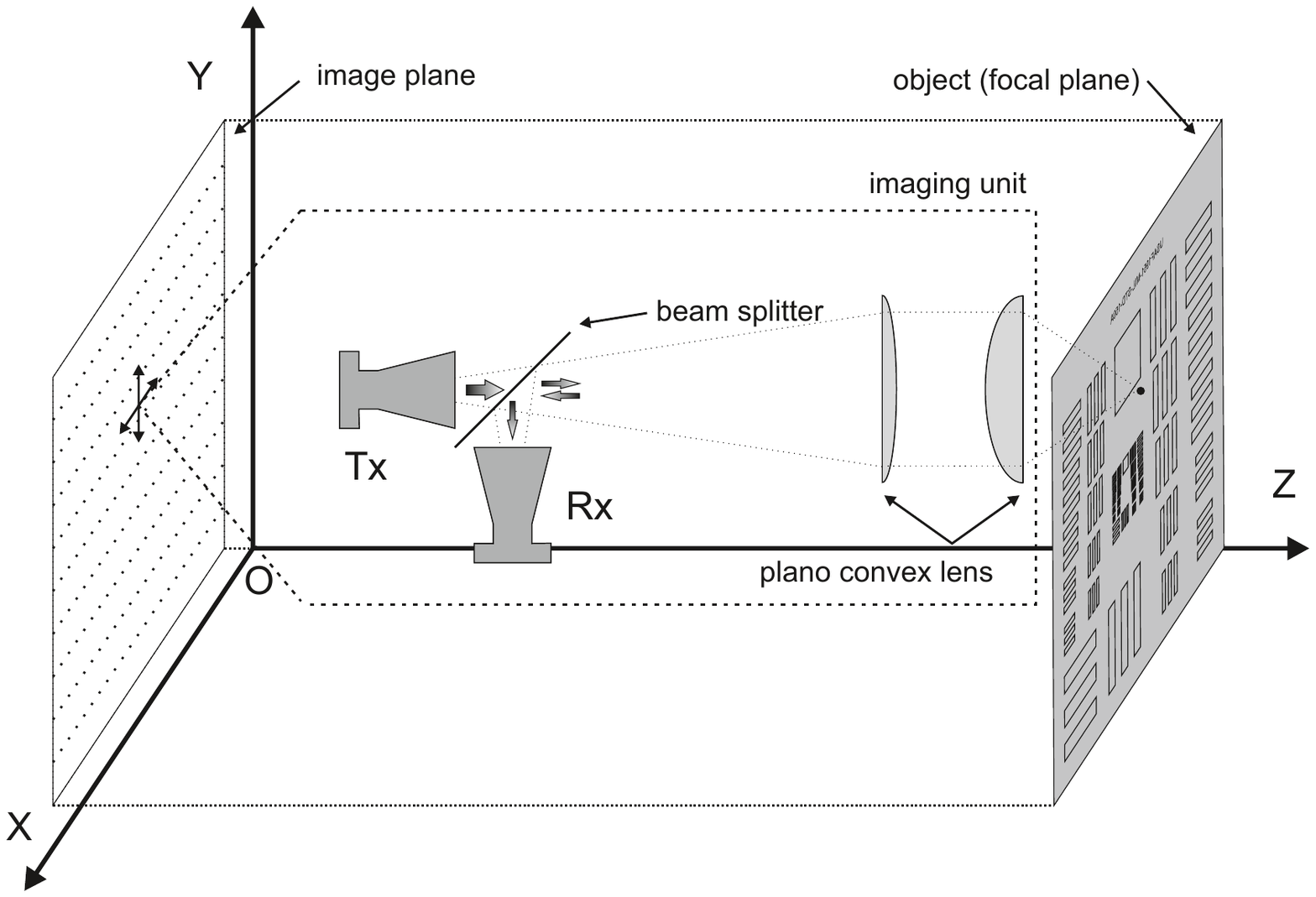}
	\includegraphics[width=0.9\textwidth]{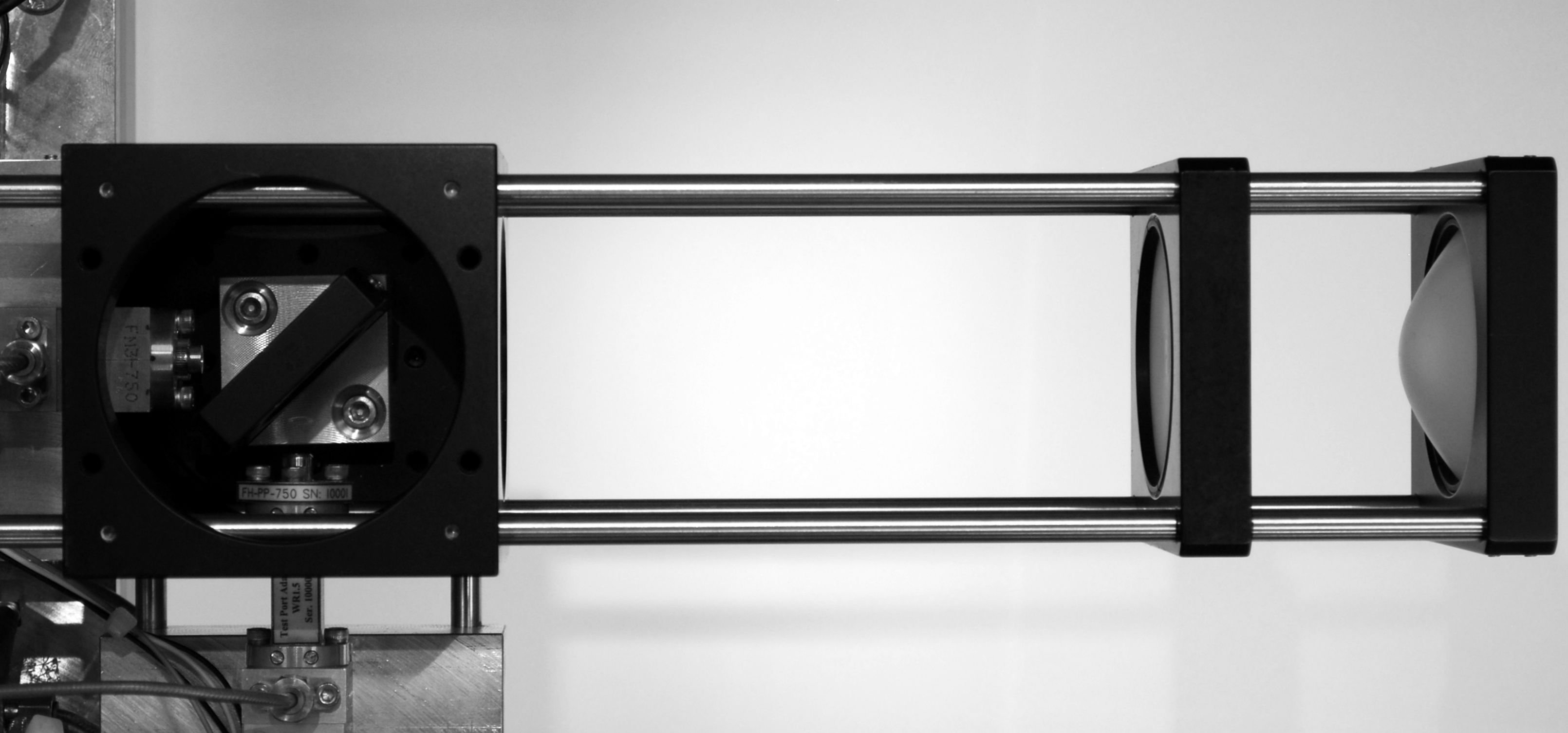}
	\caption{THz 3D imaging geometry (top) and photograph of the THz 3D imaging unit (bottom)}
	\label{fig:Geometry_new}
\end{figure}

\begin{figure}[!t]
  \centering
  \includegraphics[width=0.9\textwidth]{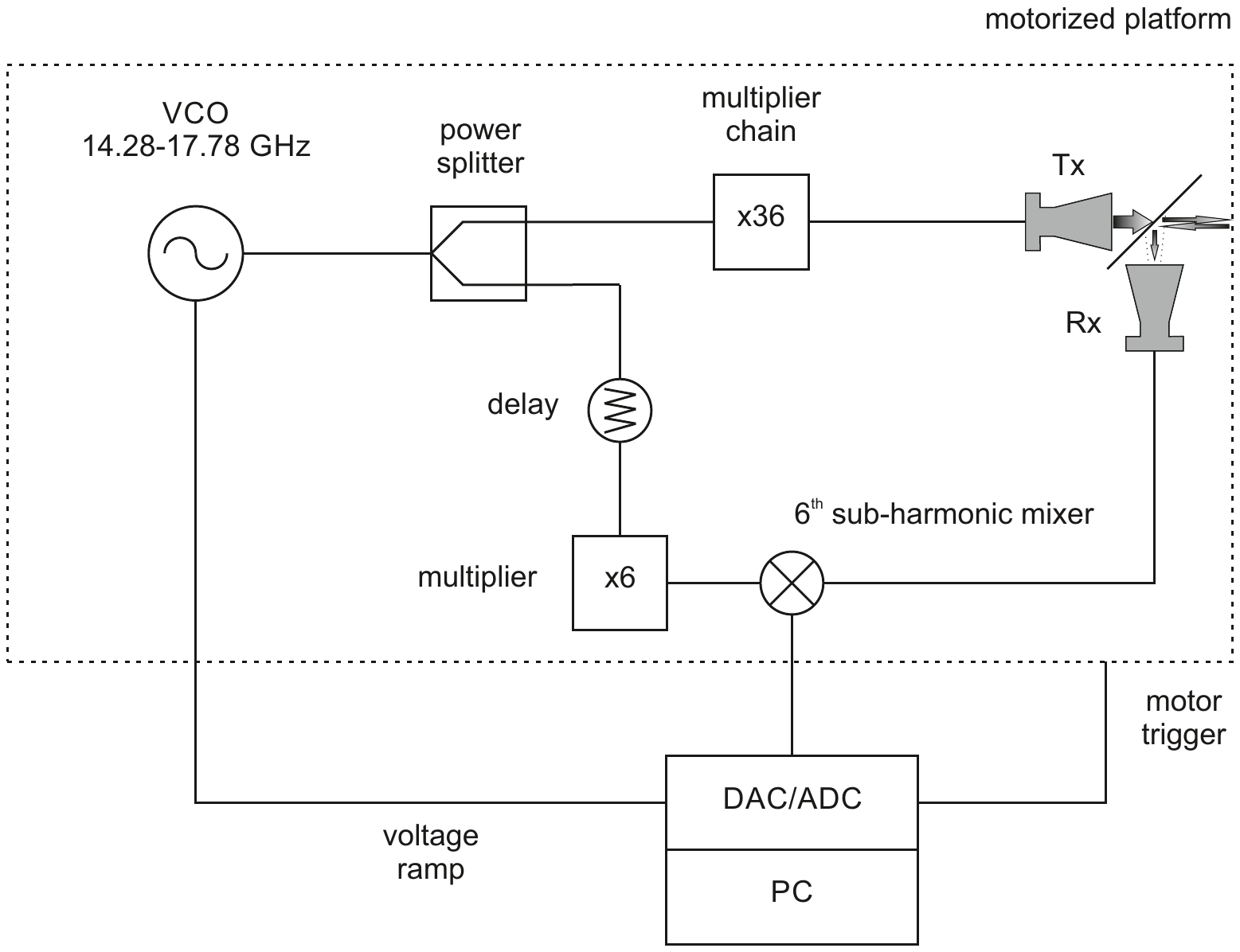}
  \caption{Schematic of the THz 3D imaging system setup and the host PC}
  \label{fig:Schematic_wo_margin}
\end{figure}

Our electronic THz imaging system is based on hollow-waveguide multipliers and mixers, operating in a frequency modulated continuous wave (FMCW) mode for measuring depth information. The components are operating around a frequency of $577$GHz with a bandwidth of $126$GHz. More details on the experimental approach are described in\cite{ding2013thz}.\par

Fig.~\ref{fig:Geometry_new} shows the imaging geometry. Both transmitter (Tx) and receiver (Rx) are mounted on the same platform. The imaging unit, consisting of Tx, Rx and optical components, are moved along the x and y direction using stepper motors and linear stages. This imaging unit takes a depth profile of the object at each lateral position, in order to acquire a full 3D image. The data is acquired with a lateral step size of $262.5\mu m$ in xy-direction. During measurement, the motor controller and the data acquisition are synchronized to enable on the fly measurements. An adequate integration time and velocity is chosen in order to provide enough time for the acquisition of 1400 samples per depth profile and 36 averages per sample. The total acquisition time for such an averaged depth profile is $5$ms.\par

Transmitter (Tx) and receiver (Rx) are mounted in a monostatic geometry. A beam
splitter and two hyperbolic lenses focus the beam radiated from the Tx
and reflected from the sample into the Rx. 
By measuring minimum dimension which obtains more than 3dB intensity difference, the resolution of the setup
is measured as $793.7\mu m$ using a metallic USAF 1951
Resolving Power Test Target scaled to the THz frequency range, which is close
to the theoretical ideal expectation of $622\mu m$
(numerical aperture NA$=0.508$~@~$578$GHz). 
Due to the monostatic scanning approach no optical magnification occurs.
The system's depth resolution $\Delta d$ is defined by
\begin{equation}\label{eqn:system_depth_resolution}
  \Delta d = \cfrac{c}{2 \times B}
\end{equation}
where $c$ is the speed of light in air, and $B$ is the system
bandwidth. For our system, we have $\Delta d=1210\mu m$\par

For the FMCW operation, a voltage controlled oscillator (VCO) is tuned from $14.28-17.78$GHz (see Fig.~\ref{fig:Schematic_wo_margin}).  The signal at the output of the VCO is distributed to the Tx and the Rx using a power-splitter. For transmission the signal is then upconverted to $514-640$GHz with a chain of multipliers. After downconversion to the intermediate frequency (IF) range using a sub-harmonic mixer, which is fed with the 6th harmonic of the VCO signal, the signal is digitized with 10 MS (mega-samples) per second sampling rate and transferred to the host PC. An additional delay between the Tx and Rx path creates a frequency offset in the intermediate frequency signal for proper data acquisition.\par


\section{Proposed method}
\label{s:method}

\subsection{Overview}
\label{s:method.method_overview}

\begin{figure}[!t]
	\centering
		\includegraphics[width=0.8\textwidth]{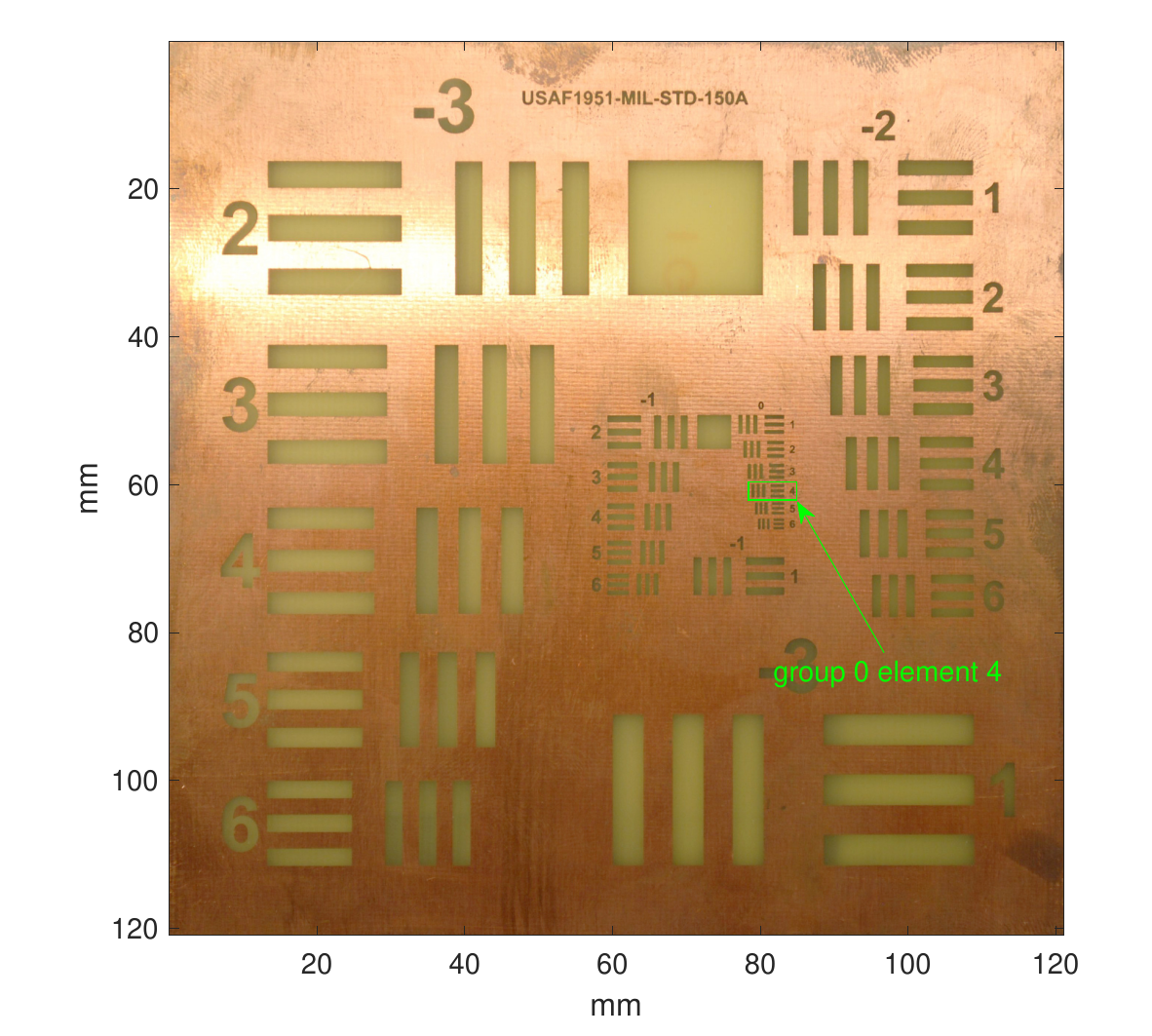}
	\caption{Metal on PCB board  test target fabricated according to USAF 1951 \emph{MIL-STD-150A} standard, where the group and row numbers indicate the lines per millimeter according to~\cite{standard1959photographic}. Group 0 Element 4 for Sec.~\ref{s:evaluation.lateral.resolution} is indicated.}
	\label{fig:usaf_photo}
\end{figure}

\begin{figure*}[!t]
	\centering
	\tikzstyle{block} = [draw, fill=white, rectangle, text width=0.18\textwidth, text centered, rounded corners, minimum height=0.05\textheight]
	\tikzstyle{joint} = [draw, circle, minimum size=0.01\textwidth]
	\tikzstyle{cloud} = [draw, fill=white, ellipse, text width=0.14\textwidth, text centered, rounded corners, minimum height=0.05\textheight]
	\tikzstyle{part} = [rounded corners, rectangle, text width=0.18\textwidth, text centered, rounded corners, minimum height=0.05\textheight , inner xsep=0.025\textwidth ,inner ysep=0.03\textheight]
	\pgfdeclarelayer{background}
	\pgfdeclarelayer{foreground}
	\pgfsetlayers{background,foreground}
	\begin{tikzpicture}[>=stealth, auto, node distance=2em]
		\begin{pgfonlayer}{foreground}
			\node [block] (peak) {Find max magnitude position};
			\node [cloud, left= 0.26\textwidth] (input) {Complex signal};
			\node [block, right=0.26\textwidth] (deconv) {Deconvolution};
			
			\node [block, below=of peak] (magnitude) {Magnitude fitting};
			\node [block, below=of magnitude] (phase) {Phase initialization};
			\node [block, below=of phase] (complex) {Complex fitting};
			\node [block, below=of complex] (recon) {Reconstruction};
			
			\node [block, below=of input] (zeropadding) {Zero-padding};
			\node [block, below=of zeropadding] (FFT) {FFT};

			\node [cloud, below=of deconv] (output) {Output Image};
			
			\draw [->] (input) -- node[right]{\(u\)}(zeropadding);
			\draw [->] (zeropadding) -- node[right]{\(u_{k}\)}(FFT);
			
			\draw [->] (FFT) --++ (2.3,0) node [near start,anchor=east,above,pos=0.38]{\(I_u,\hat{u}\)} |- (peak);
			
			\draw [->] (peak) -- node[right]{\(z_{\text{max}}\)}(magnitude);
			\draw [->] (magnitude) -- node[right]{\(A_{m},\mu_{m},\sigma_{m}\)}(phase);
			\draw [->] (phase) -- node[right]{\(A_{m},\mu_{m},\sigma_{m}, \phi_{m}\)}(complex);
			
			\draw [->] (complex) -- node[right]{\(A,\mu,\sigma,\phi\)}(recon);
			\draw [->] (recon) --++ (2.3,0) node[near start,above,pos=0.38]{\(d_v,I_v\)} |- (deconv);
			\draw [->] (deconv) -- node[right]{\(I_d\)}(output);
		\end{pgfonlayer}
			
		\begin{pgfonlayer}{background}
			\node [part,fill=blue!20,fit=(input)(zeropadding)(FFT),label={Preprocessing}](pp){};
			\node [part,fill=green!20,fit=(peak)(magnitude)(complex)(recon),label={Parameter extraction}](cf){};
			\node [part,fill=red!20,fit=(deconv)(output),label={Deconvolution}](dc){};
		\end{pgfonlayer}
	\end{tikzpicture}
	\caption{Block diagram of proposed method (see intensity images in Fig.~\ref{fig:intensity_comparison1} and Fig.~\ref{fig:intensity_comparison2})}
	\label{fig:block_diagram}
\end{figure*}
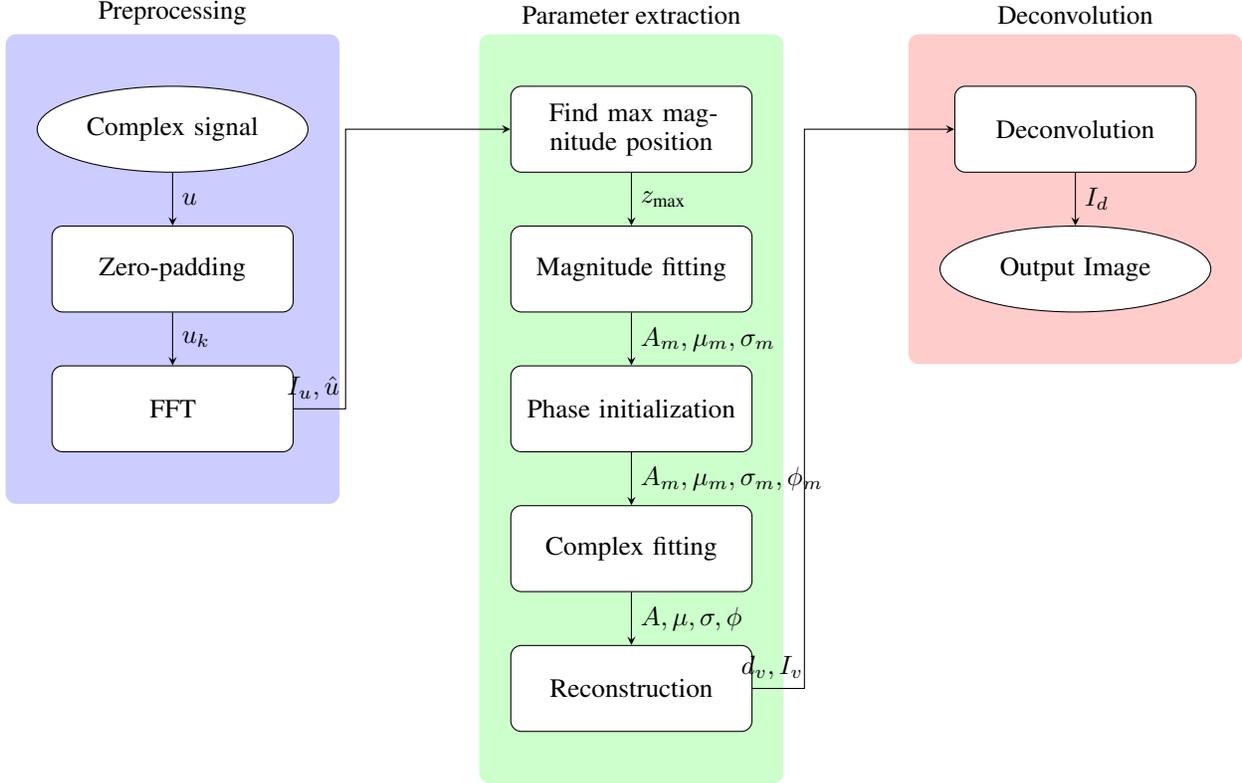

In the previous THz 3D imaging system~\cite{ding2013thz}, the signal was assumed to have an ideal flat target with perfect orthogonal alignment to the THz sensor.
However, perfect planarity and orthogonality require a high precision of the manufacturing procedure and calibration of the acquisition setup. To study more realistic THz imaging scenarios, we allow for \emph{non-planar} targets (see Figs.~\ref{fig:usaf_photo} and \ref{fig:stepchart_photo}), which are not perfectly orthogonally aligned to the sensor, \ie that the distance between sensor to a lateral pixel in xy-direction is an unknown variable.\par

Fig.~\ref{fig:block_diagram} depicts our method, which comprises three major components: pre-processing, parameter extraction and deconvolution. 
In the pre-processing part, the measured complex signal is interpolated by zero-padding to obtain more sampling points in $z$-direction.  
In the parameter extraction part, a complex model is fitted to the in-phase and quadrature components of the signal in z-direction for each lateral position. 
From this fitting, we deduce corrected reflectance complex field signal and depth information. 
In the deconvolution part, we process the reconstructed 2D image with deconvolution algorithm to form a high resolution image in xy-domain.\par

\subsection{Preprocessing}
\label{s:method.preprocessing}

With the initial input data to our computational procedure, we have a complex THz signal acquired per-pixel with frequency ramping~\cite{munson1989signal}. 
In this paper, $u(x,y)$ is denoted as the measured complex signal at lateral position $(x,y)$ with discrete frequency parameter $k$ with length $N_z$.\par

In order to achieve sub-wavelength geometric correction, more sampling points on the z-axis are required for robust curve fitting. 
Based on the current acquisition system (see Sec.~\ref{s:system}), an intuitive
method is to interpolate the signal in the spatial domain, but the time
domain signal provides another option. Instead of spatial interpolation, we
extend the discrete signal by a factor of $N$ using zero-padding
of the complex electric field signal $u(x,y)[k]$ in the frequency domain:
\begin{equation}\label{eqn:zero_padding}
  u_{N}(x,y)[k] = \begin{cases} u(x,y)[k], & \mbox{if } k < N_z
    \\ 0, & \mbox{otherwise} \end{cases},
\end{equation}
where $N$ is the zero-padding factor, and the length of $u_{N}$ is
$D = N\cdot N_z$. In this paper, we use $N = 9$.\par

After zero-padding, the signal is transformed into the spatial domain by
applying a deramp-FFT~\cite{munson1989signal}.
\begin{equation}\label{eqn:fourier_tran}
  \hat{u}(x,y) = \mathcal{F}\{u_{N}(x,y)\}.
\end{equation}
The resulting 3D image \(\hat{u}\) can be expressed as a 3D matrix
in the spatial xyz-domain, representing per-pixel $(x,y)$ the complex reflectivity
of THz energy in z-direction represented by the complex samples
$\hat{u}(x,y)[z_0]\to \hat{u}(x,y)[z_{D-1}]$.\par


\section{Parameter extraction}
\label{s:parameter_extraction}

In this part, we apply per-pixel parameter extraction in z-direction in
order to represent the measured complex signal by a complex
reflection model. As each pixel is treated independently, we simplify
notation by dropping the pixel-location using, \eg, $\hat u[z_{i}]$
for $\hat u(x,y)[z_{i}]$.\par

Our FMCW-THz 3D imaging system is calibrated by amplitude normalization with respect to an ideal metallic reflector. 
Thus, we achieve an ideal rectangular frequency amplitude signal response, which, after being
Fourier transformed, results in an ideal $\sinc$ function $v(z)$ as
continuous spatial signal amplitude. 
In this model, we assume single layer reflection, and the extension to multi-layer reflection is part of future work.
As we assume a single layer reflection, the complex signal model $v(z)$ is modeled as a modulated $\sinc$ function:
\begin{equation}\label{eqn:model_complex}
\begin{split}
  & v(z) = A \cdot \sinc(\sigma(z-\mu)) \cdot e^{-j(\omega z-\phi)} \\[0.5em]
  \mbox{where, } & \sinc(t) = \begin{cases} \cfrac{\sin(\pi t)}{\pi t} & t \neq 0 \\
    1 & t = 0 
  \end{cases}	
\end{split}
\end{equation}
In Eq.~\eqref{eqn:model_complex}, $A$ is the electric field amplitude,
\(\mu\) and \(\sigma\) are the mean (\ie the depth) and the width of the $\sinc$ function, respectively, \(\omega\) is the angular frequency of the sinusoidal carrier and \(\phi\) is the depth-related phase shift. 
We formulate the parameter extraction as a complex curve fitting process with respect to optimizing and minimizing the energy function \(\varepsilon_{c}\)
\begin{equation}\label{eqn:energy_complex}
\begin{split}
  \varepsilon_{c} = \arg\min_{A,\mu,\sigma,\phi} \sum_{z \in Z_{f}}^{}
  & \left[ \left( \hat{u}_{re}[z_{i}] - v_{re}(z|A,\mu,\sigma,\phi) \right)^2 \right. \\
  & \left. + \left( \hat{u}_{im}[z_{i}] - v_{im}(z|A,\mu,\sigma,\phi) \right)^2 \right]\\
\end{split}
\end{equation}
where the subscripts ${re}$ and ${im}$ denote the real and the imaginary part of a complex number, respectively, 
and $Z_f$ is the fitting window (see Sec.~\ref{s:parameter_extraction.window}).\par

Because of the highly non-linear optimization involved in the curve fitting process, 
a direct application of Eq.~\eqref{eqn:energy_complex} to a non-linear solver potentially results in local minima and does not lead to robust results. 
Therefore, we apply the following optimization steps which achieve a robust complex curve fitting:
\begin{enumerate}
\item Estimate the signal's maximum magnitude z-position $z_m$ in
  order to localize the \emph{curve fitting window} (see
  Sec.~\ref{s:parameter_extraction.window}).
\item Apply a \emph{curve fitting to the magnitude signal} leading
  to initial values for $A_m,\mu_m,\sigma_m$ (see
  Sec.~\ref{s:parameter_extraction.magnitude-fitting}).
\item Estimate the initial phase value $\phi_m$ using a
  \emph{phase matching} with respect to the angle of complex signal $\angle\hat{u}$ (see
  Sec.~\ref{s:parameter_extraction.phase}).
\item Based on the initial values $A_{m}, \mu_{m}, \sigma_{m}, \phi_m$
  the optimization is performed by minimizing the energy  $\varepsilon_{c}$ (see Eq.~\eqref{eqn:energy_complex}) using the Trust Region  Algorithm~\cite{coleman1996interior}.
\item Reconstruct an intensity image $I_v$ and an depth image $d_v$ based on the curve fitting result (sec Sec.~\ref{s:parameter_extraction.reconstruction}).
\end{enumerate}

Our curve fitting approach significantly reduces the per-pixel intensity inhomogeneity (see Sec.~\ref{s:evaluation.depth.cross}). 
The subsequent lateral deconvolution algorithm discussed in Sec.~\ref{s:parameter_extraction.deconvolution} involves the numerical solution of an ill-posed inverse problem of finding the blur kernel and enhancing the image's sharpness at the same time, which is very sensitive to noise. 
Therefore, correcting the intensity yields two advantages, i.e., it stabilizes the numerical deconvolution process and it prevents wrong interpretations of intensity variation as structural or material transitions. 
We will provide comparison in Fig.~\ref{fig:intensity_comparison2} and discuss the lateral resolution enhancement with and without the per-pixel parameter extraction in Sec.~\ref{s:evaluation.lateral.resolution}.

\subsection{Curve Fitting Window}
\label{s:parameter_extraction.window}

As we focus on the primary reflection signal, assuming that the
geometric reflection energy concentrates on the first air-material
interface, we locate the z-position that exhibits the maximum magnitude within a  fitting window
\begin{equation}\label{eqn:window}
  Z_{f} \in [ z_{\text{max}} - \tau_{f} , z_{\text{max}} + \tau_{f} ]
\end{equation}
with center 
\begin{equation}\label{eqn:z_peak}
  z_{\text{max}} = \arg\max_{z_{i}} \left|\hat{u}[z_{i}]\right|,
\end{equation}
\ie, \(z_{m}\) is the maximum magnitude z-position in the complex
spatial domain data.  \(\tau_{f}\) is the half-width of the fitting
window.  The choice of \(\tau_{f}\) is discussed in
Sec.~\ref{s:evaluation.optimization}.

\subsection{Magnitude Curve Fitting}
\label{s:parameter_extraction.magnitude-fitting}
Since the complex model \(v(z)\) in Eq.~\eqref{eqn:model_complex} is non-linear and the optimization for $\varepsilon_c$ in Eq.~\eqref{eqn:energy_complex} is non-convex,
the estimation of the initial parameters is critical in order to avoid local minima. A reliable initial estimate of the complex curve fitting parameters \(A_{m},\mu_{m},\sigma_{m}\) is deduced from a
magnitude curve fitting.\par

The magnitude signal model \(v_{m}(z)\) is
derived from Eq.~\eqref{eqn:model_complex} and is expressed as
\begin{equation}\label{eqn:mag_gauss}
  v_{m}(z) = A_{m} \cdot \left|\sinc\left(\sigma_{m}(z-\mu_{m})\right)\right|,
\end{equation}
where \(A_{m}\) is the electric field amplitude based on signal
magnitude, \(\mu_{m}\) is the center of $\sinc$ function, and
\(\sigma_{m}\) is the width. The magnitude curve fitting minimizes
the energy function \(\varepsilon_{m}\) by the Trust-Region Algorithm~\cite{coleman1996interior}
\begin{equation}\label{eqn:energy_magnitude_fitting}
  \varepsilon_{m} = \arg\min_{A_{m},\mu_{m},\sigma_{m}} \sum_{z \in Z_{f}}^{} \left( |\hat{u}[z_{i}]| - v_{m}(z|A_{m},\mu_{m},\sigma_{m}) \right)^{2}.
\end{equation}

After the magnitude curve fitting, $A_m,\mu_m,\sigma_m$ are the
initial values for $A,\mu,\sigma$ with respect to the complex signal model in
Eq.~\eqref{eqn:model_complex}. 
However, an estimate for the phase angle $\phi_m$ is still required.

\subsection{Estimating the Initial Phase Value \texorpdfstring{\(\phi_m\)}{}}
\label{s:parameter_extraction.phase}
We assume that, within the fitting window $Z_f$, 
the phase angle $\omega z-\phi$ in our model (see Eq.~\eqref{eqn:model_complex}) matches the phase angle $\angle\hat{u}$ in the spatial domain data $\hat u[z_i]$.
The corresponding optimization energy functional measures the linear deviation between these phase angles as follows
\begin{equation}\label{eqn:phase_init_energy}
\begin{split}
  \varepsilon_{\phi} = \arg\min_{\phi_m} \cfrac{1}{2} \sum_{z \in Z_f}{} & \left[ \left( \cos(\phi_m-\omega z)-\cos(\angle\hat{u}) \right)^2 \right. \\
    & \left. + \left( \sin(\phi_m-\omega z)-\sin(\angle\hat{u}) \right)^2 \right]
\end{split}
\end{equation}
Setting the gradient of $\varepsilon_{\phi}$ to zero and applying trigonometric identities, we obtain 
\begin{equation}\label{eqn:phase_init_identity}
\sin\phi_m \sum_{z \in Z_f}{\cos(\omega z+\angle\hat{u})} = \cos\phi_m \sum_{z \in Z_f}{\sin(\omega z+\angle\hat{u})}
\end{equation}
In Eq.~\eqref{eqn:phase_init_identity}, the initial phase angle $\phi_m$ is independent from the data angle $\angle\hat{u}$. 
Therefore, the minimum to Eq.~\eqref{eqn:phase_init_energy} is found by solving for $\phi_m$ in Eq.~\eqref{eqn:phase_init_identity}, yielding
\begin{equation}\label{eqn:phase_init_calculation}
\phi_m = \tan^{-1} \cfrac{\sum_{z \in Z_f}{\sin(\omega z+\angle\hat{u})}} {\sum_{z \in Z_f}{\cos(\omega z+\angle\hat{u})}}
\end{equation}

After this phase initialization, $A_m,\mu_m,\sigma_m$ and $\phi_m$ are
given as initial values for $A,\mu,\sigma$ and $\phi$ in the model in
Eq.~\eqref{eqn:model_complex}, respectively, and the model is fitted
according to the energy function $\varepsilon_c$ in
Eq.~\eqref{eqn:energy_complex} using the Trust Region
Algorithm~\cite{coleman1996interior}.

After the complex curve fitting, the four different parameters of our model in Eq.~\eqref{eqn:model_complex} are extracted: amplitude $A$, mean $\mu$, width $\sigma$, and phase \(\phi\).
Because of scattering and multi-layer reflection, error exists if we fit in an ideal sinc-function. Therefore, $\sigma$ is extracted as a varying parameter to indicate the error. The depth parameter $\mu$ is evalauted in Sec.~\ref{s:evaluation.depth}. The amplitude parameter $A$ is further processed in a 2D approach. The processing method on all other parameters ($\mu$, $\sigma$, $\phi$) will be investigated in future research.

\subsection{Intensity Image Reconstruction}
\label{s:parameter_extraction.reconstruction}

Next, we have to extract the per-pixel intensities using our model. We, therefore, first define the reference intensity image $I_u$ based on the input data  $\hat u$ as the intensity of the z-slice with the maximum average magnitude:
\begin{equation}\label{eqn:source_intensity}
\begin{split}
  & I_{u}(x,y) = \hat u[z_{mean}] \cdot \hat u[z_{mean}]^* \\[0.5em]
  \mbox{where, } & z_{mean} = \arg\max_{z_i} \cfrac{\sum_{x,y}{\left|\hat{u}(x,y)[z_{i}]\right|}}{N_x N_y}
\end{split}
\end{equation}
where \(\hat{u}^*\) is the complex conjugate of \(\hat{u}\) and
$N_x,N_y$ are the size of the matrix in x-axis and y-axis, respectively.
The \emph{reconstructed intensity} is deduced from the curve fitted
data model and is defined as the intensity of the model's signal
(Eq.~\eqref{eqn:model_complex}) at the center position $\mu$:
\begin{equation}\label{eqn:output_intensity_1}
  \begin{split}
    I_v(x,y) & = v(x,y,\mu) \cdot v(x,y,\mu)^* \\
    & = A^2(x,y) \cdot sinc^2(0) \\
    & = A^2(x,y)
  \end{split}
\end{equation}

\section{Deconvolution}
\label{s:parameter_extraction.deconvolution}

After the curve fitting, the reconstructed intensity \(I_v\) is a 2D image with real and positive values. We can now apply a state-of-art 2D deconvolution algorithm to enhance the xy-domain resolution. In contrast to prior work we use a total variation (TV) blind deconvolution algorithm to improve the spatial resolution~\cite{xu2010two,xu2013unnatural}. \par

In deconvolution image model, our reconstructed intensity image \(I_v\) can be expressed as a blurred observation of a sharp image \(I_d\)
\begin{equation}\label{eqn:convolution_model}
  I_v = I_d \ast h + \eta,
\end{equation}
where $h$ is the \emph{spatially invariant} point spread function
(PSF) (also known as \emph{blur kernel}) and $\eta$ is the noise;
$\ast$ denotes convolution. Blind deconvolution methods allow for
estimating the blur kernel directly from the data, which is, however,
an ill-posed inverse problem that requires a
prior knowledge in order to deduce a robust result~\cite{levin2011understanding}. In this paper, we
utilize the sparse nature of intensity
gradients~\cite{perrone2014total} and choose a state-of-art TV blind
deconvolution algorithm that minimizes
\begin{equation}\label{eqn:deconvolution_model}
  \left(I_d,h\right) = \arg\min_{I_d,h} \|I_d \ast h - I_v\|_1 + \lambda \| \nabla  I_d \|_1.
\end{equation}
Here, $\|I_d \ast h - I_v\|_1$ is commonly referred as a \emph{data term}, $\lambda$ is a \emph{regularization parameter} and $\| \nabla  I_d \|_1$ is the \emph{TV-regularization} (or \emph{prior}) that enforces the gradient of the resulting deblurred image $I_d$ to be sparse. As by-product, the blind deconvolution yields the estimated PSF \(h\). We obtain our final results using  the implementation from Xu\etal\cite{xu2010two,xu2013unnatural}.  \par



\section{Evaluation}
\label{s:evaluation}

\begin{figure}[!t]
  \centering
  \includegraphics[width=0.7\textwidth]{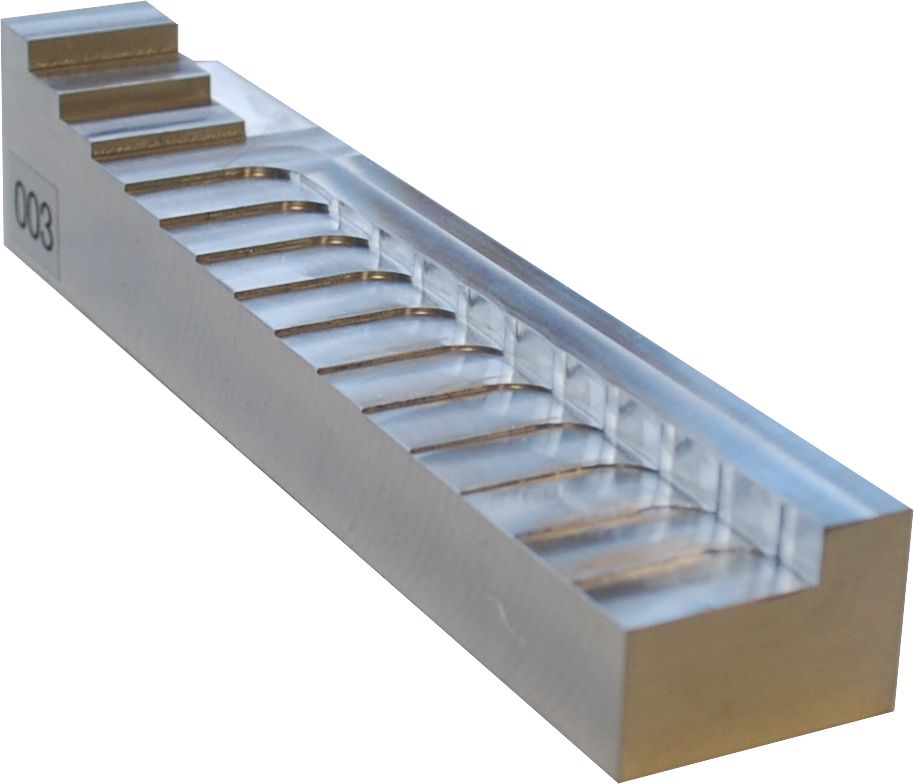}
  \caption{Metallic step object with a reference zero}
  \label{fig:stepchart_photo}
\end{figure}

In this section we evaluate our image enhancement approach using the
following two datasets
\begin{enumerate}
\item \textbf{MetalPCB}: A nearly planar ``USAF'' target etched on a
  circuit board (Fig.~\ref{fig:usaf_photo}). The dataset has been
  acquired using the setup described in Sec.~\ref{s:system} and has
  the resolution $N_x=446$, $N_y=446$, $N_z=1400$.  The lateral
  per-pixel distance (\ie the sensor mechanical movement distance) is
  $262.5{\mu}m$. 
  The target is etched in the standard size scale of USAF target \emph{MIL-STD-150A}.
\item \textbf{StepChart}: A metallic step chart with steps varying
  from $4000\mu m$ to $50\mu m$, and a reference
  plane to locate the reference zero position
  (Fig.~\ref{fig:stepchart_photo}). The dataset has also been acquired
  using the setup in Sec.~\ref{s:system} with the resolution
  $N_x=575$, $N_y=113$, $N_z=1400$.  The lateral per-pixel distance is
  $262.5{\mu}m$.
\end{enumerate}
It should be noted, that our system has a lateral resolution of $622\mu m$ as its ideal diffraction limit,
an experimentally measured lateral resolution $793.7\mu m$ and a depth accuracy of $\Delta d=1210\mu m$ as its diffraction limits (see Sec.~\ref{s:system}).

We apply and compare the following images, reconstruction and image enhancement methods.
The images are named as $\langle image \rangle \_ \langle method \rangle\mbox{-}\langle kernel \rangle$, 
where $image$ is the intensity image applying deconvolution methods, 
$method$ is the method applied to the intensity image, 
and $kernel$ is the deblur kernel adopted in the deconvolution method.
\begin{enumerate}
\item \ReferNoIE: The reference intensity image $I_u$; no reconstruction using curve fitting and no image enhancement applied.
\item \ReconstNoIE: The reconstructed intensity image $I_v$ using curve fitting; no image enhancement applied.
\item \ReferXu: image enhancement using Xu\etal\cite{xu2010two, xu2013unnatural} applied to reference intensity \ReferNoIE.
\item \ReconstXu: image enhancement using Xu\etal\cite{xu2010two, xu2013unnatural} applied to reconstructed intensity \ReconstNoIE. This is our proposed method.
\item \ReferLRG: using Lucy-Richardson~\cite{lucy1974iterative, richardson1972bayesian} with gaussian kernel applied to reference intensity \ReferNoIE.
\item \ReconstLRG: using Lucy-Richardson~\cite{lucy1974iterative, richardson1972bayesian} with gaussian kernel applied to reconstructed intensity \ReconstNoIE.
\item \ReferLRXu: using Lucy-Richardson~\cite{lucy1974iterative, richardson1972bayesian} with sparse kernel extracted from Xu\etal\cite{xu2010two, xu2013unnatural} applied to reference intensity \ReferNoIE.
\item \ReconstLRXu: using Lucy-Richardson~\cite{lucy1974iterative, richardson1972bayesian} with sparse kernel extracted from Xu\etal\cite{xu2010two, xu2013unnatural} applied to reconstructed intensity \ReconstNoIE.
\end{enumerate}

In the following, we will first deduce the optimal window size for the quality control of the fitting using the \textbf{MetalPCB} dataset
(Sec.~\ref{s:evaluation.optimization}). 
In Sec.~\ref{s:evaluation.depth}, we will discuss the depth accuracy using the \textbf{StepChart} dataset.
In Sec.~\ref{s:evaluation.lateral}, we will discuss the lateral resolution on \textbf{MetalPCB} dataset. 
All intensity images are normalized to a perfect metal reflection, and displayed using MATLAB's  perceptionally uniform colormap \textit{parula}~\cite{matlab2016}. \par

\subsection{Window Size Optimization}
\label{s:evaluation.optimization}

\begin{figure}[!t]
  \centering
  \includegraphics[width=0.9\textwidth]{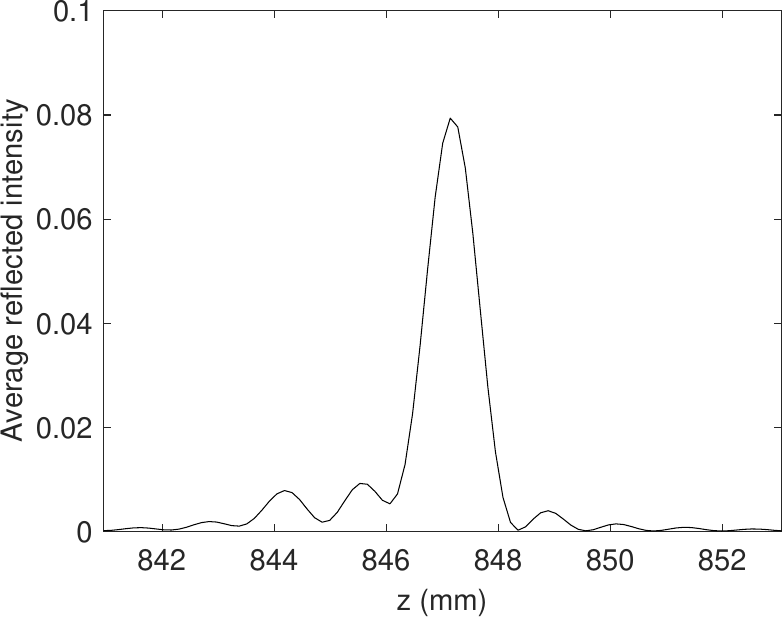}
  \caption{Average reflected intensity of $\hat{u}$ in PCB region.}
  \label{fig:average_intensity}
\end{figure}

\begin{figure}[!t]
  \centering
  \includegraphics[width=0.9\textwidth]{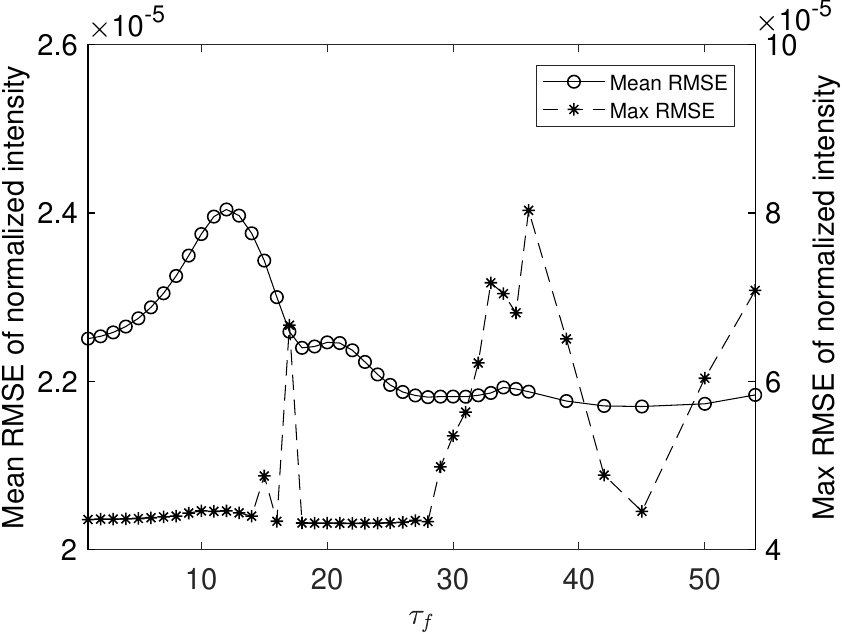}
  \caption{Mean RMSE and maximum RMSE by fitting window width $\tau_f$.}
  \label{fig:rmse_plot}
\end{figure}

In Fig.~\ref{fig:average_intensity}, the average intensity of
$\hat{u}$ by the z-axis in the PCB region of the  \textbf{MetalPCB} dataset is shown. 
We can observe that, compared to the symmetric model in Eq.~\eqref{eqn:model_complex}, 
the z-axis signal has a lower main lobe to side lobe ratio in the PCB region. 
This might be due to the superposition of signal reflection from the front and back PCB surfaces (see Fig.~\ref{fig:usaf_photo}). 
This indicates that a large
fitting window size $\tau_{f}$ in Eq.~\eqref{eqn:window} would corrupt
any quantitative evaluation and should be avoided. On the other hand, a
small fitting window size is also not feasible, as we need a
sufficient number of sampling points to get a robust fitting result and to avoid over-fitting.

To obtain a reliable numeric measurement, we evaluate the fitting error for varying fitting window half-widths $\tau_f$ using the Root-Mean-Square-Error (RMSE) between the fitted model $v$ (see Eq.~\eqref{eqn:model_complex}) and the spatial data $\hat{u}$:
\begin{equation}\label{eqn:rmse_complex_fitting}
  \mbox{RMSE}(x,y) = \sqrt{ \cfrac{\sum_{z \in Z_{f}}^{} |\hat{u}(x,y)[z]-v(x,y,z)|^2}{2\tau_{f}+1} }
\end{equation}
As a measure for the full THz image, we calculate the mean and the maximum RMSE over all pixels
\begin{equation}\label{eqn:rmse_complex_fitting_2}
  \begin{split} 
    \mbox{Mean RMSE} & = \cfrac{\sum_{x,y}^{} \mbox{RMSE}(x,y)}{N_x N_y} \\  
    \mbox{Max RMSE} & = \max_{x,y} (\mbox{RMSE}(x,y)) \\
  \end{split}
\end{equation}

In Fig~\ref{fig:rmse_plot}, the mean and the maximum RMSE of the curve fitting with a different fitting window $\tau_{f}$ are shown. 
We can see that the mean RMSE and maximum RMSE are both increasing when $\tau_f \leq 13$, which is expected due to over-fitting. 
When the fitting window is increased to a larger value, we can observe that the mean RMSE is decreased steadily until $\tau_f=45$. 
The maximum RMSE, however, has no clear tendency and strongly varies  beyond $\tau_f=28$. 
After considering that the mean and maximum RMSE are both considerably low when the window size is 45, we choose $\tau_f=45$ as our reference and optimal window size, which is used throughout the rest of the evaluation.\par
  
\subsection{Depth Accuracy and Intensity Reconstruction}
\label{s:evaluation.depth}

In this section, we analyzed the depth accuracy enhancement in the z-direction obtained by the proposed curve fitting method. 
In Sec.~\ref{s:evaluation.depth.accuracy}, the depth accuracy of the \textbf{StepChart} dataset using the maximum magnitude method and the proposed method are shown. 
In Sec.~\ref{s:evaluation.depth.cross}, we analyze the variance of a homogeneous metal cross section on the \textbf{MetalPCB} dataset.\par

\subsubsection{Depth Accuracy}
\label{s:evaluation.depth.accuracy}

\begin{figure}[!t]
  \centering
  \includegraphics[width=0.9\textwidth]{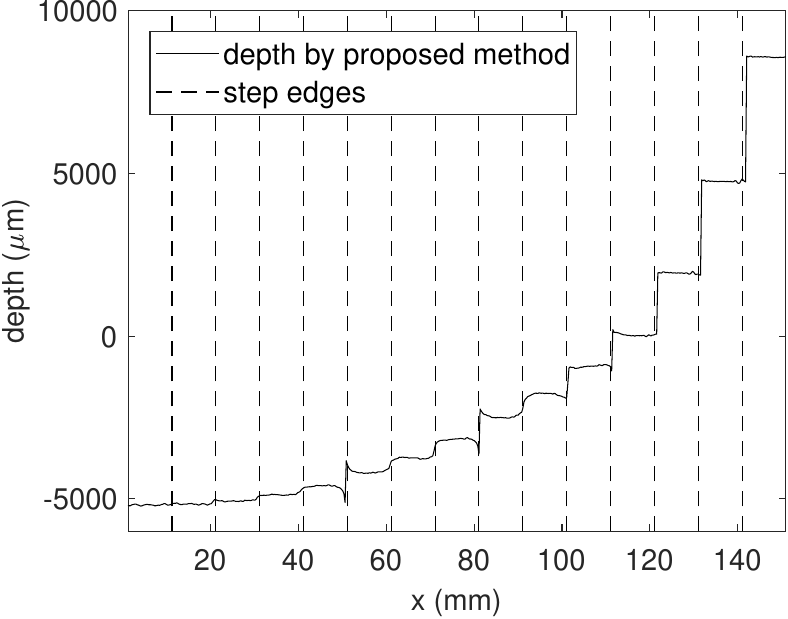}
  \caption{Cross section depth of \textbf{StepChart} dataset by the proposed method and the step edges}
  \label{fig:depth_plot}
\end{figure}

\begin{table}[!t]
\renewcommand{\arraystretch}{1.3}
\centering
\caption{Depth difference, depth difference standard deviation (SD) and error comparison between maximum magnitude and the proposed curve fitting method}
\label{tab:depth_resolution}
\begin{tabular}{r|rr|rr|rr}
\toprule
\multicolumn{5}{c|}{Depth Difference} & \multicolumn{2}{c}{Relative Error} \\ \hline
\multicolumn{1}{c|}{$\mbox{depth}_{gt}$} & \multicolumn{2}{c|}{$\mbox{depth}_{\text{max}}$} & \multicolumn{2}{c|}{$\mbox{depth}_{\mu}$} & $\mbox{error}_{\text{max}}$ & $\mbox{error}_{\mu}$ \\
 & mean & SD & mean & SD & & \\
($\mu m$) & ($\mu m$) & ($\mu m$) & ($\mu m$) & ($\mu m$) & ($\%$) & ($\%$) \\ \hline
 4009.0 &  3898.5 &     7.2 &  3831.3 &    12.4 &    2.76 &    4.43 \\ 
 2987.0 &  2797.2 &    54.0 &  2810.8 &    13.0 &    6.35 &    5.90 \\ 
 2006.0 &  1908.7 &    53.6 &  1926.8 &    19.4 &    4.85 &    3.95 \\ 
 1004.0 &   941.1 &     0.0 &   958.6 &    23.4 &    6.26 &    4.53 \\ 
  903.0 &   806.7 &     0.0 &   815.3 &    17.5 &   10.67 &    9.71 \\ 
  803.0 &   792.8 &    40.9 &   742.0 &    14.0 &    1.27 &    7.60 \\ 
  703.0 &   633.8 &    77.3 &   665.5 &    20.5 &    9.84 &    5.33 \\ 
  600.0 &   590.0 &    65.6 &   561.9 &    19.5 &    1.66 &    6.35 \\ 
  472.0 &   403.3 &     0.0 &   468.8 &    13.9 &   14.55 &    0.68 \\ 
  410.0 &   403.3 &     0.0 &   391.0 &    13.7 &    1.63 &    4.64 \\ 
  298.0 &   268.9 &     0.0 &   287.6 &    14.7 &    9.77 &    3.48 \\ 
  208.0 &   268.9 &     0.0 &   192.2 &    15.1 &  -29.27 &    7.60 \\ 
   91.0 &    17.7 &    45.5 &    89.3 &    17.4 &   80.58 &    1.88 \\ 
   42.0 &   102.2 &    61.8 &    34.9 &    21.9 & -143.28 &   16.84 \\ 
\bottomrule
\end{tabular}
\end{table}

In this part, we evaluate the depth accuracy using the \textbf{StepChart} dataset. 
In comparison to the depth of the proposed method $\mbox{depth}_{\mu}$, we obtain the depth $\mbox{depth}_{\text{max}}$ using the maximum magnitude position $z_{\text{max}}$ (Eq.\eqref{eqn:z_peak}) of $\hat{u}$.\par

The z-positions are both calibrated to $\mu m$ by the reference zero z-position $z_0$
\begin{equation}
  \label{eqn:depth_normalization}
  \begin{split}
    \mbox{depth}_{\mu} & = \cfrac{\mu-z_0}{N} \cdot \Delta d\\[0.5em]
    \mbox{depth}_{\text{max}} & = \cfrac{z_{\text{max}}-z_0}{N} \cdot \Delta d
  \end{split}
\end{equation}
where $N$ is the zero-padding factor in Eq.\eqref{eqn:zero_padding}, $\Delta d=1210\mu m$ is the physical depth per sample, \ie the systems depth resolution (see Eq.~\eqref{eqn:system_depth_resolution}).\par

Fig.~\ref{fig:depth_plot}, the cross section depth of \textbf{StepChart} dataset is plotted with an expected position of the edges. 
We can observe an interference effect due to signal superposition at several edges, most notably at  $x=50$mm and $x=81$mm. 
Even though blind deconvolution can hardly resolve strong interference effects, spatially varying point-spread-function is required in order to cope with these kind of effects in the  deconvolution stage; see also Hunsche\etal\cite{hunsche1998thz}.\par 

In order to circumvent interference effects, we extract and average depth values from the center 350 samples for each step. 
Then, we calculate the depth differences between adjacent steps. 
These depth differences are compared to the ground truth values, which is obtained by mechanical measurement.
Tab.~\ref{tab:depth_resolution} depicts the depth differences and the corresponding standard deviation (SD) of the ground truth $\mbox{depth}_{gt}$, the maximum magnitude method $\mbox{depth}_{\text{max}}$ and the proposed curve fitting $\mbox{depth}_{\mu}$ (see Eq.~\eqref{eqn:depth_normalization}). In order to compare the depth accuracy, we calculate the relative error as
\begin{equation}
  \label{eqn:depth_error}
  \begin{split}
    \mbox{error}_{\mu} & = \cfrac{\mbox{depth}_{\mu} - \mbox{depth}_{gt}}{\mbox{depth}_{gt}}\\[0.5em]
    \mbox{error}_{\text{max}} & = \cfrac{\mbox{depth}_{\text{max}} - \mbox{depth}_{gt}}{\mbox{depth}_{gt}}
  \end{split}
\end{equation} 
In this paper, we consider a depth difference as resolvable when the relative error is below $10\%$ with a reasonably low deviation.
Thus, our proposed method can still resolve the $91\mu m$ depth difference, while the maximum magnitude method can only resolve depth difference up to $298\mu m$. 
As a result, the proposed curve fitting method enhances the system depth accuracy to $91\mu m$.\par

\subsubsection{Intensity Reconstruction}
\label{s:evaluation.depth.cross}

\begin{figure}[!t]
	\centering
	\subfloat[\label{fig:intensity_u}]{\includegraphics[width=0.5\textwidth]{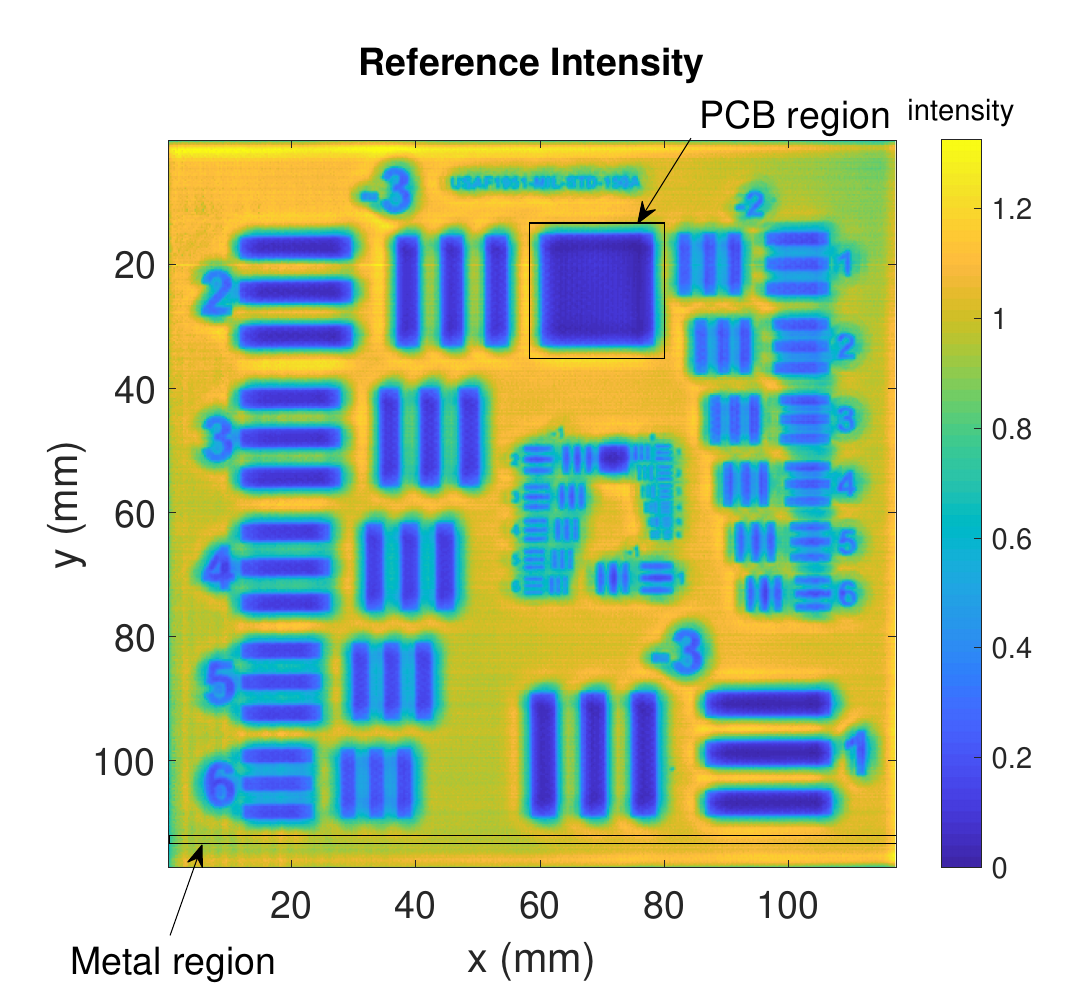}}
	\hfill
	\subfloat[\label{fig:intensity_v}]{\includegraphics[width=0.5\textwidth]{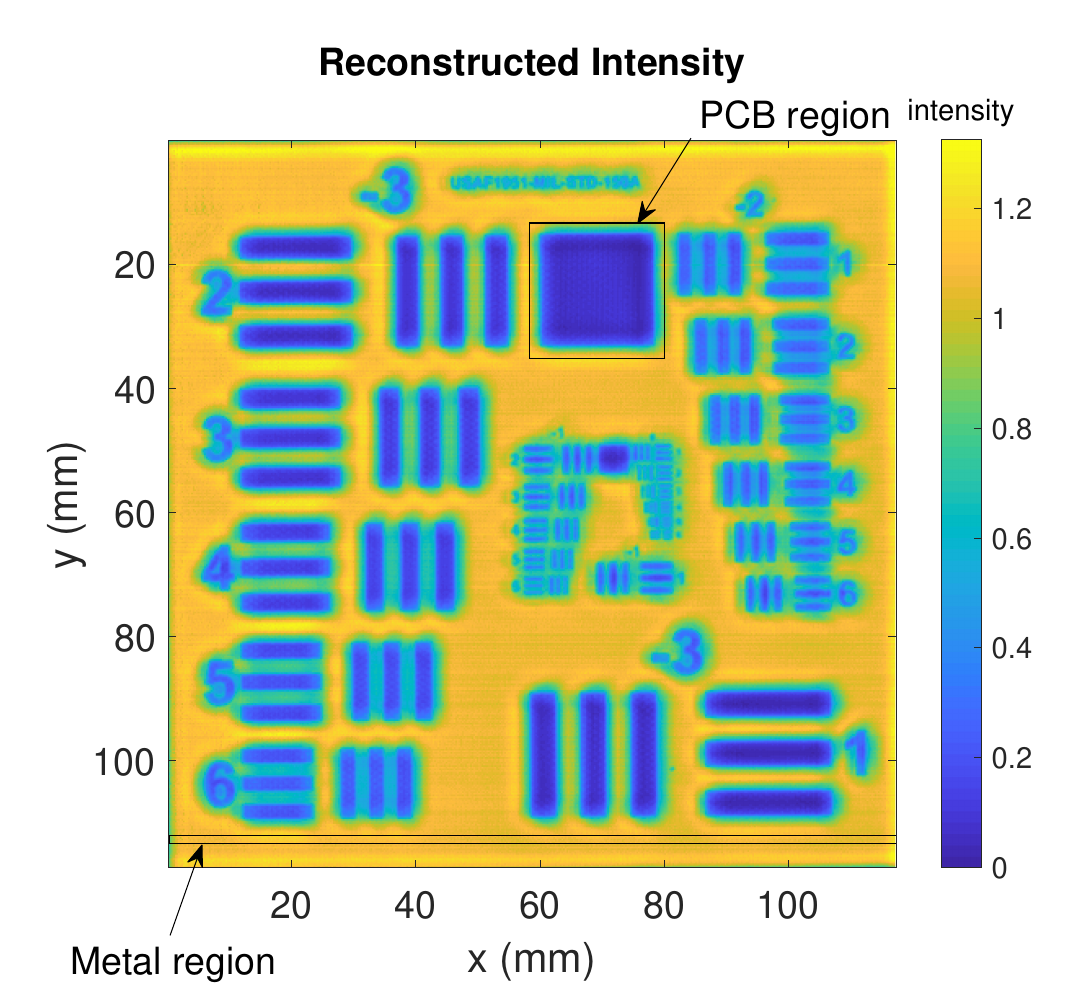}}
	\hfill
	\caption{Comparison between
	\protect\subref{fig:intensity_u} Reference intensity $I_u$ (\ReferNoIE)
	\protect\subref{fig:intensity_v} Reconstructed intensity $I_v$ (\ReconstNoIE). The homogeneous metal regions for Sec.~\ref{s:evaluation.depth.cross} and the PCB region for Sec.~\ref{s:evaluation.lateral.cross} are indicated.}
	\label{fig:intensity_comparison1}
\end{figure}

\begin{figure}[!t]
  \centering
  \includegraphics[width=0.9\textwidth]{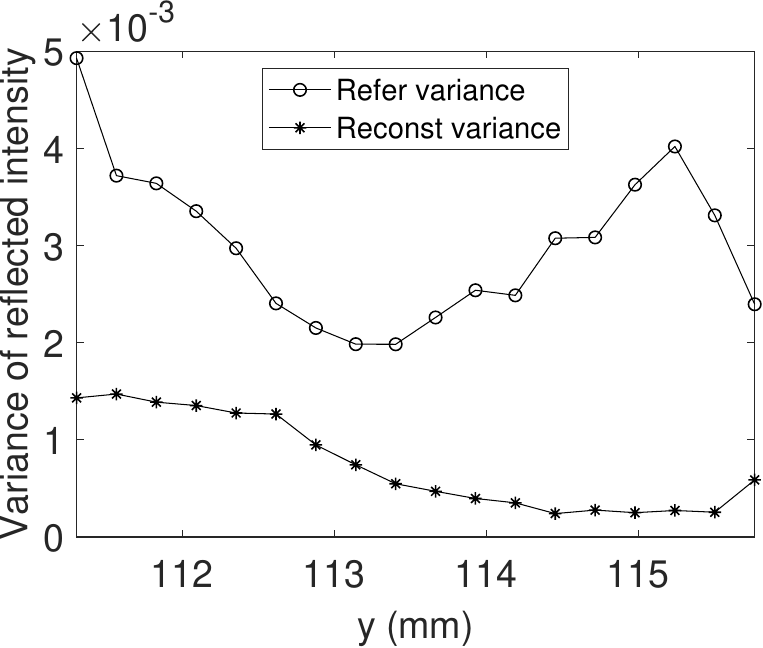}
  \caption{Evaluation of the variance of the reflected
      intensities along the homogeneous metal regions in
      Fig.~\ref{fig:intensity_u} and Fig.~\ref{fig:intensity_v} in
      $y$-direction: For each pixel-line (parameterized over the
      $y$-position in mm) we compute the variance of the reference
      intensity $I_u$ (\ReferNoIE) and the reconstructed
      intensity $I_v$ (\ReconstNoIE) is given, showing the
      improvement resulting from the curve fitting in
      Sec.~\ref{s:parameter_extraction}.}
  \label{fig:homogeneous_intensity_variance}
\end{figure}

In this part we evaluate the reconstructed intensity image that is directly deduced from our enhanced depth accuracy according to Eq.~\eqref{eqn:output_intensity_1}. We, therefore, use the \textbf{MetalPCB} dataset that comprises large homogeneous copper regions. The \textbf{MetalPCB} target is, however, not perfectly flat and/or not perfectly aligned orthogonally. \par

Fig.~\ref{fig:intensity_comparison1} depicts the intensity images
for the reference intensity $I_u$ based on
Eq.~\eqref{eqn:source_intensity} (Fig.~\ref{fig:intensity_u}) and the
proposed reconstructed intensity $I_v$ using
Eq.~\eqref{eqn:output_intensity_1} (Fig.~\ref{fig:intensity_v}).  We
can observe that the copper regions do not appear to be fully
homogeneous in reference intensity image $I_u$. After applying the high accuracy depth
reconstruction, this intensity inhomogeneity is significantly reduced
in the reconstructed image $I_v$.\par

Fig.~\ref{fig:homogeneous_intensity_variance} further analyzes the
intensity homogeneity using a horizontal metal region (see
Fig.~\ref{fig:intensity_comparison1}, lower image part). The
horizontal metal region consists of 18 pixel rows for each of which
we compute the intensity variance in the reference intensity image
$I_u$ and the reconstructed intensity image $I_v$. We observe that the
reconstruct intensity $I_u$ has a significantly reduced intensity
variance within all rows of copper, as the reconstruct intensity is
focused on an accurate depth position for each lateral position.\par

Both, the visual and the numerical results in
Fig.~\ref{fig:intensity_comparison1} and
Fig.~\ref{fig:homogeneous_intensity_variance}, respectively,
demonstrate that our proposed intensity reconstruction method achieves
significantly improved homogeneous intensities on homogeneous material
regions without introducing additional distortions or noise.\par 

\subsection{Lateral Resolution and Analysis}
\label{s:evaluation.lateral}

In this section, we analyzed the lateral domain enhancement by the proposed method using the \textbf{MetalPCB} dataset. 
In Sec.~\ref{s:evaluation.lateral.resolution}, we compared the proposed method to Lucy-Richardson's deconvolution method on behalf of the lateral resolution and in Sec.~\ref{s:evaluation.lateral.cross} we visually analyze the silk structure embedded in the PCB region.\par

\subsubsection{Lateral resolution}
\label{s:evaluation.lateral.resolution}

\begin{figure}[!t]
	\centering
	\subfloat[\label{fig:intensity_contrast_horizontal}]{\includegraphics[width=0.49\textwidth]{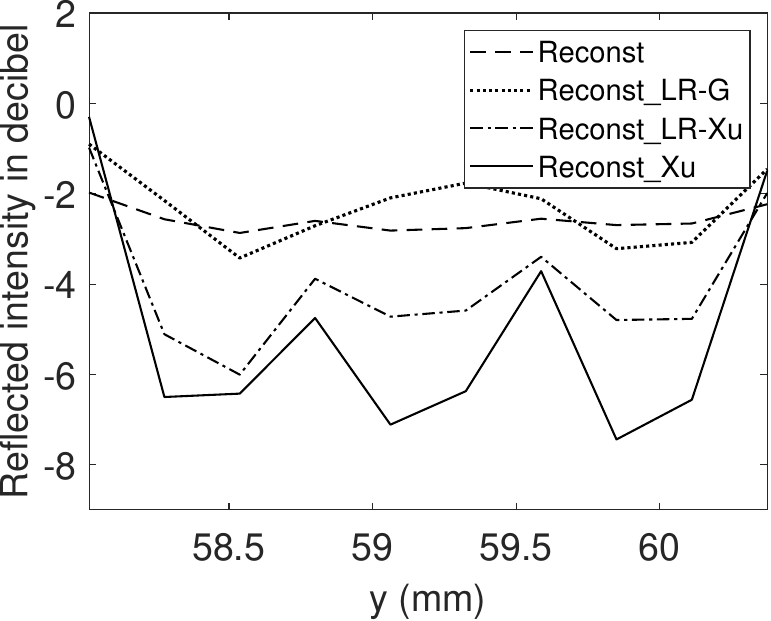}}
	\hfill
	\subfloat[\label{fig:intensity_contrast_vertical}]{\includegraphics[width=0.49\textwidth]{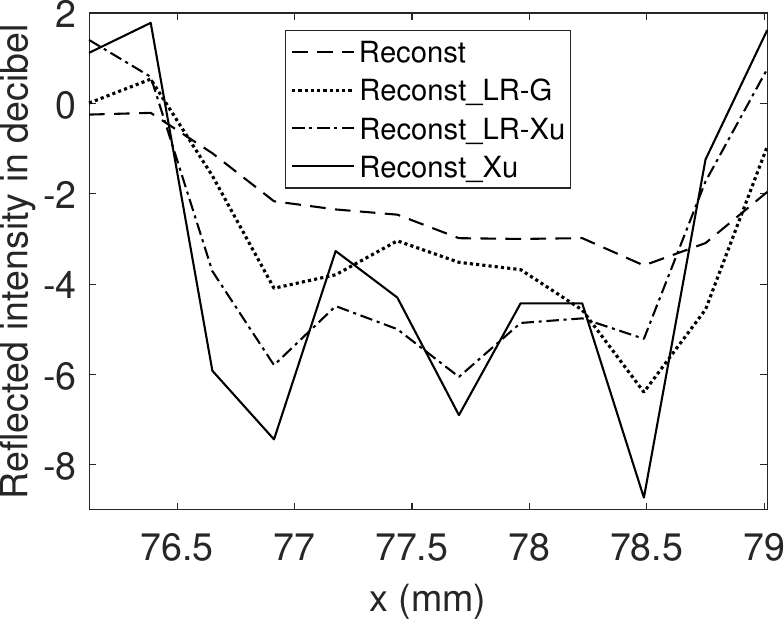}}
	\hfill
  	\caption{Cross section intensity comparison at Group 0 Element 4 (see Fig.~\ref{fig:usaf_photo}) for different deconvolution methods. This element represents a $353.6\mu m$ line distance in
  	\protect\subref{fig:intensity_contrast_horizontal} vertical direction
  	\protect\subref{fig:intensity_contrast_vertical} horizontal direction}
  	\label{fig:intensity_contrast}
\end{figure}

\begin{figure*}[!t]
	\centering
	\subfloat[\label{fig:kernel_gauss}]{\includegraphics[width=0.5\textwidth]{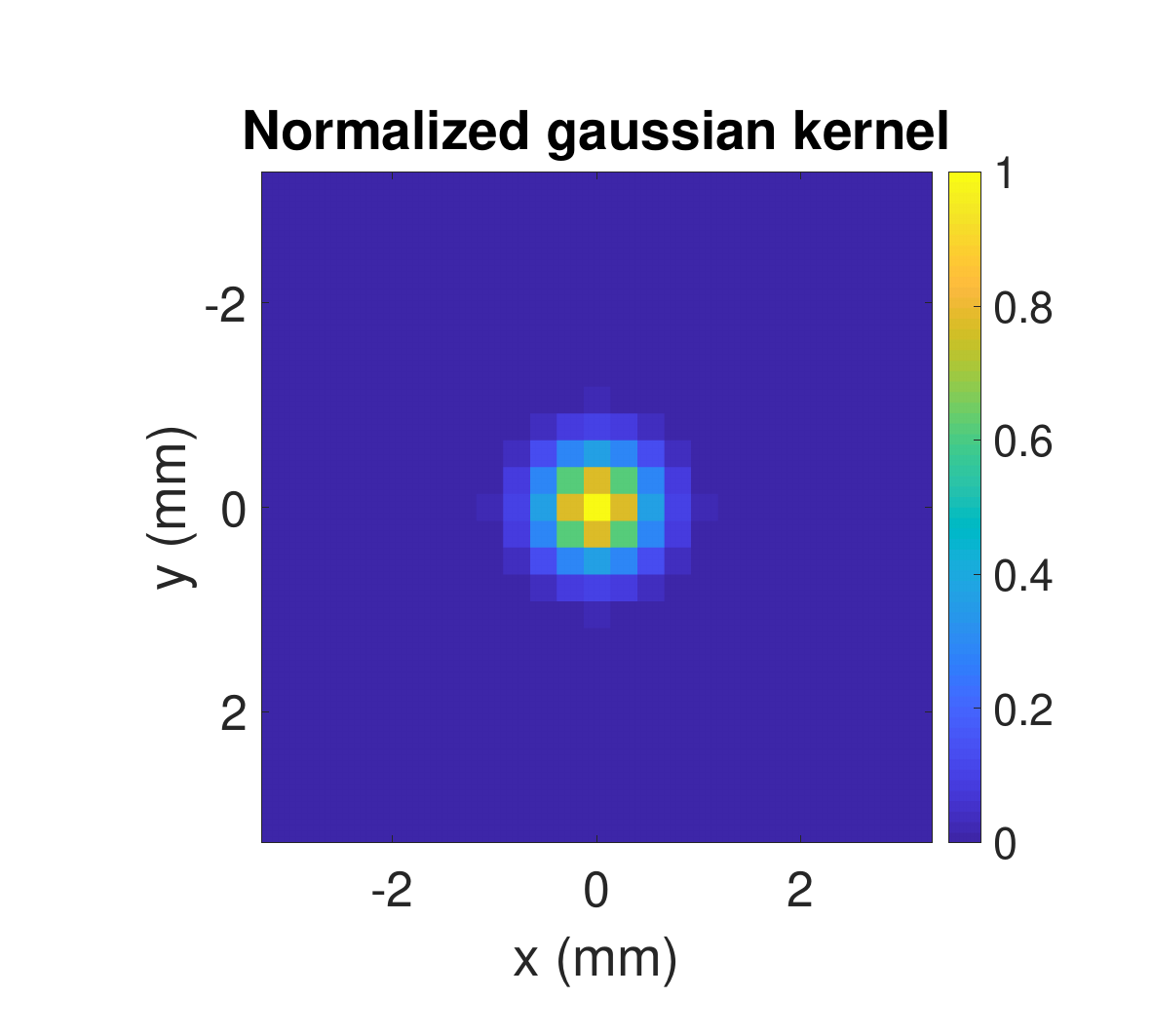}}
	\hfill
	\subfloat[\label{fig:kernel_sparse}]{\includegraphics[width=0.5\textwidth]{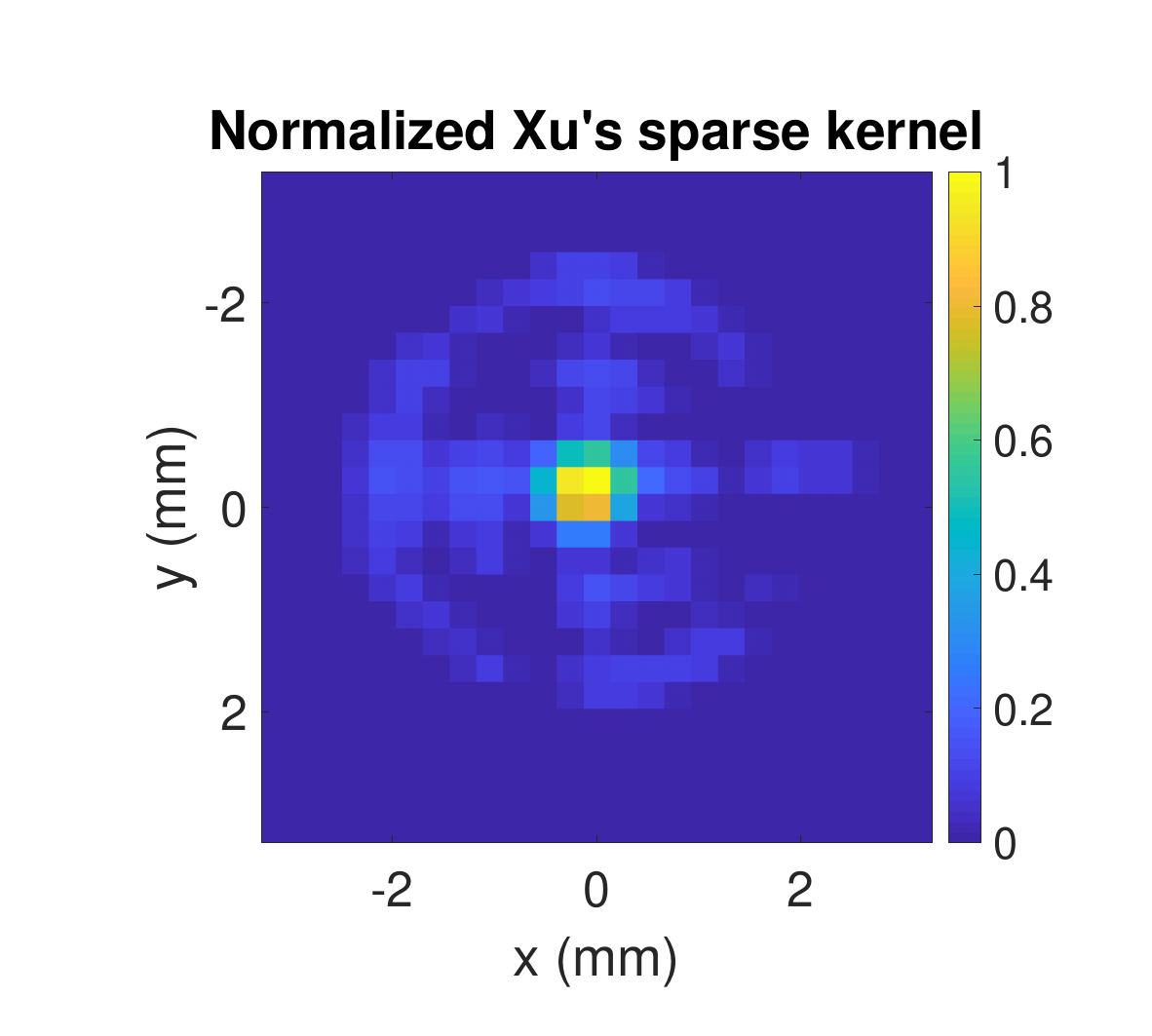}}
	\hfill
	\subfloat[\label{fig:kernel_cross}]{\includegraphics[width=0.6\textwidth]{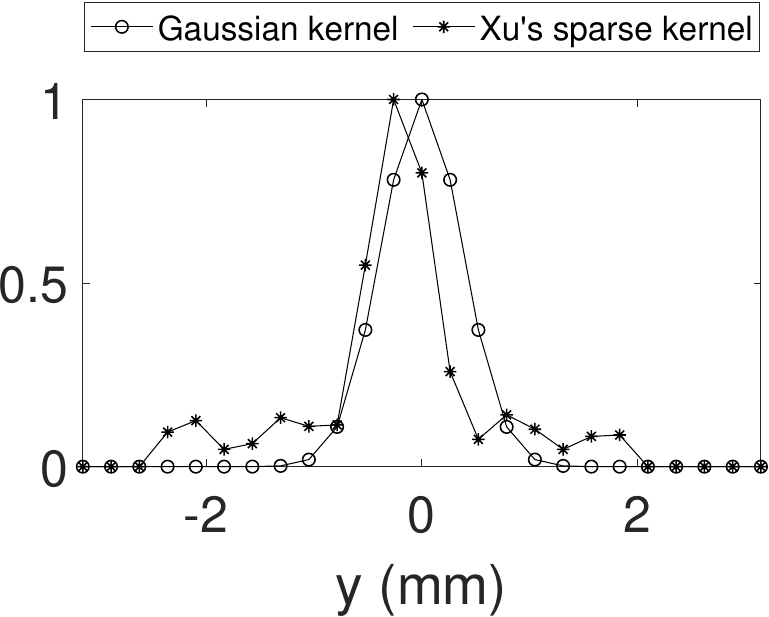}}
	\hfill
	\caption{Comparison between
	\protect\subref{fig:kernel_gauss} gaussian kernel
	\protect\subref{fig:kernel_sparse} Xu's sparse kernel
	\protect\subref{fig:kernel_cross} cross section of gaussian and Xu's sparse kernel}
	\label{fig:kernel_comparison}
\end{figure*}

\begin{figure*}[!p]
  \centering
 
  \subfloat[\label{fig:intensity_u_lg}\ReferLRG (LR, gaussian; refer. int. $I_u$)]
           {\includegraphics[width=0.39\textwidth]{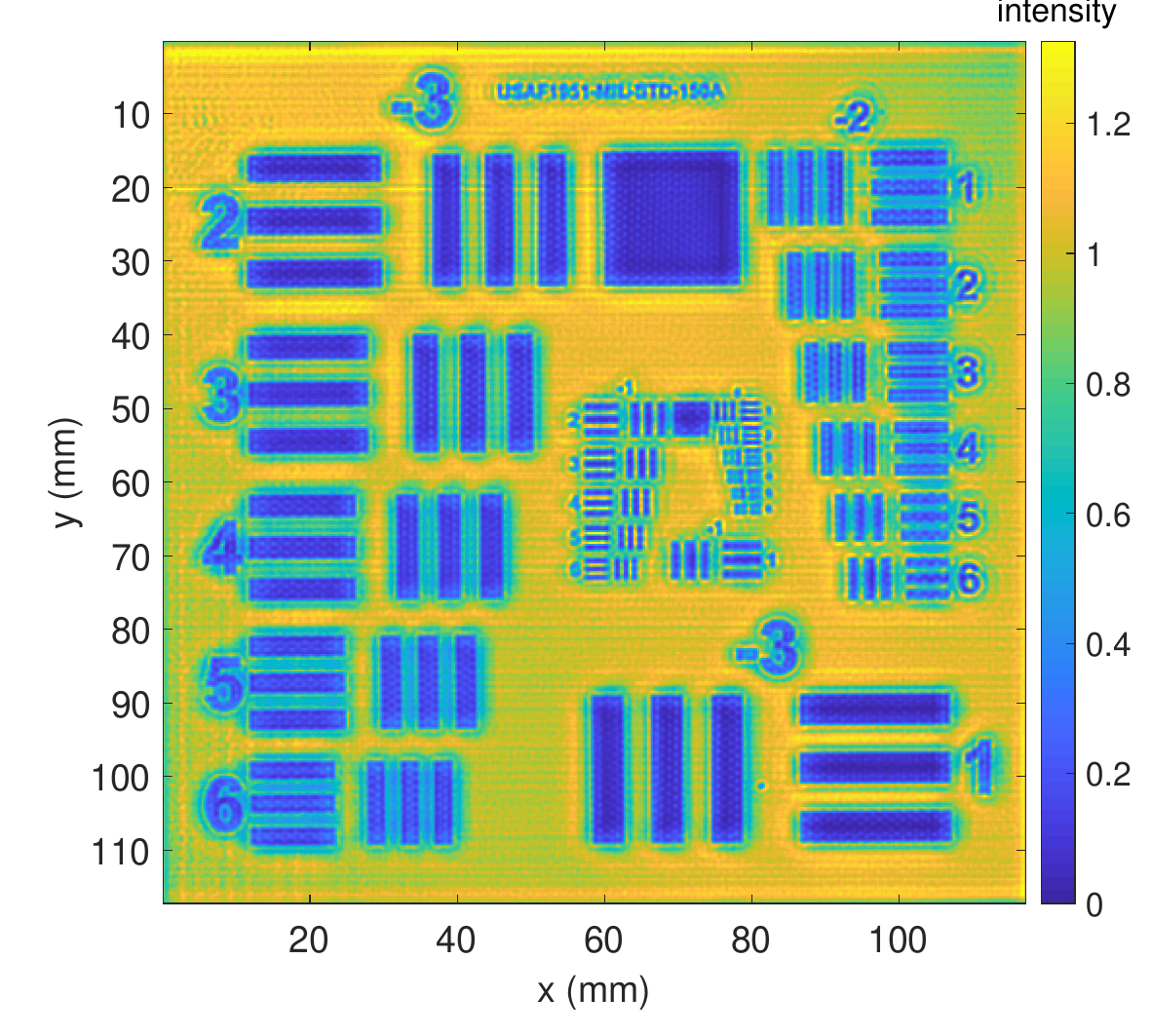}}
           \hspace{1cm}
  \subfloat[\label{fig:intensity_v_lg}\ReconstLRG (LR, gaussian; reconst. int. $I_v$)]
           {\includegraphics[width=0.39\textwidth]{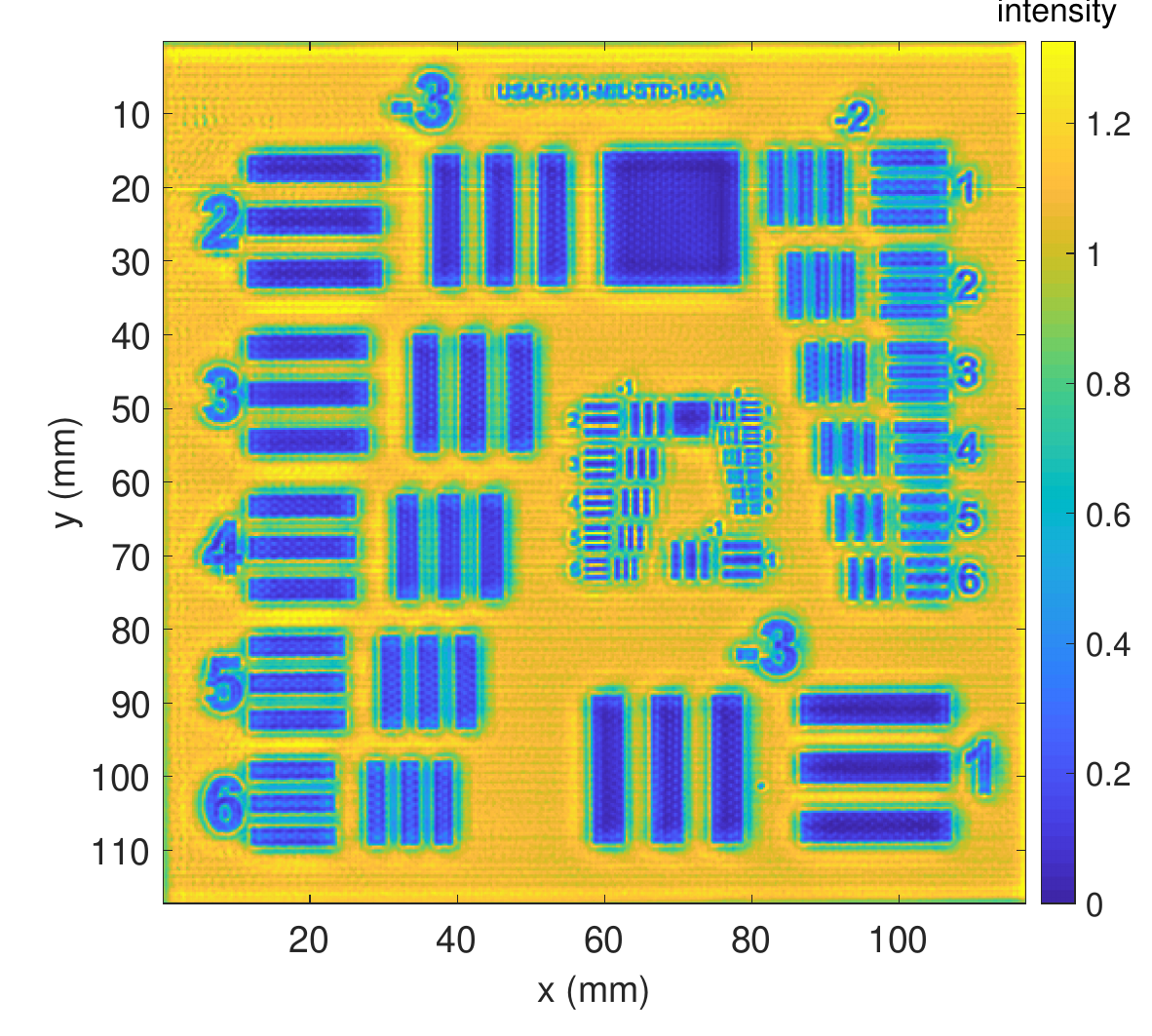}}
           
  \hfill
  \vfill		

  \subfloat[\label{fig:intensity_u_lx}\ReferLRXu (LR, Xu's kernel; refer. int. $I_u$)]
           {\includegraphics[width=0.39\textwidth]{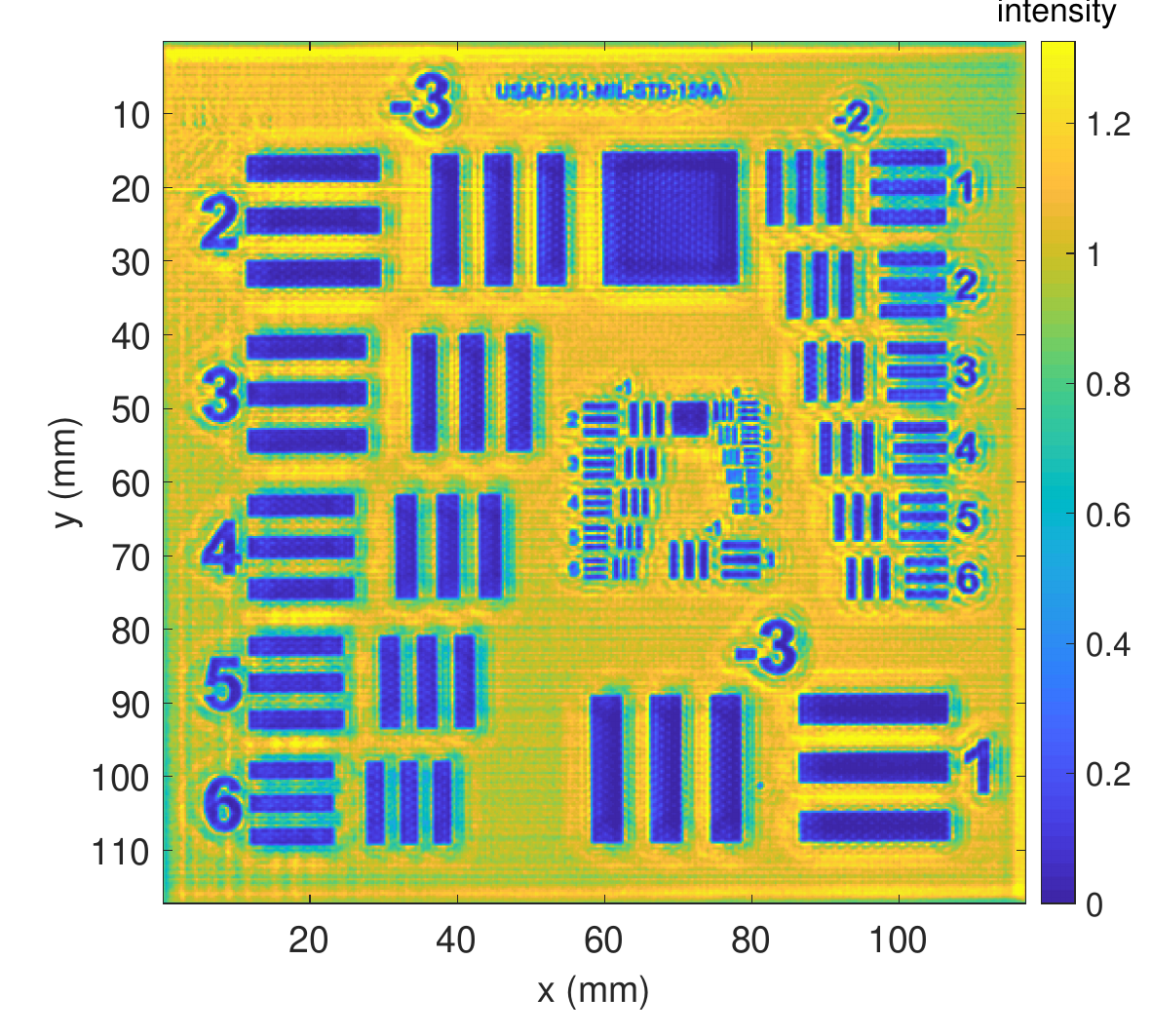}}
           \hspace{1cm}
  \subfloat[\label{fig:intensity_v_lx}\ReconstLRXu (LR, Xu's kernel; reconst int. $I_v$)]
           {\includegraphics[width=0.39\textwidth]{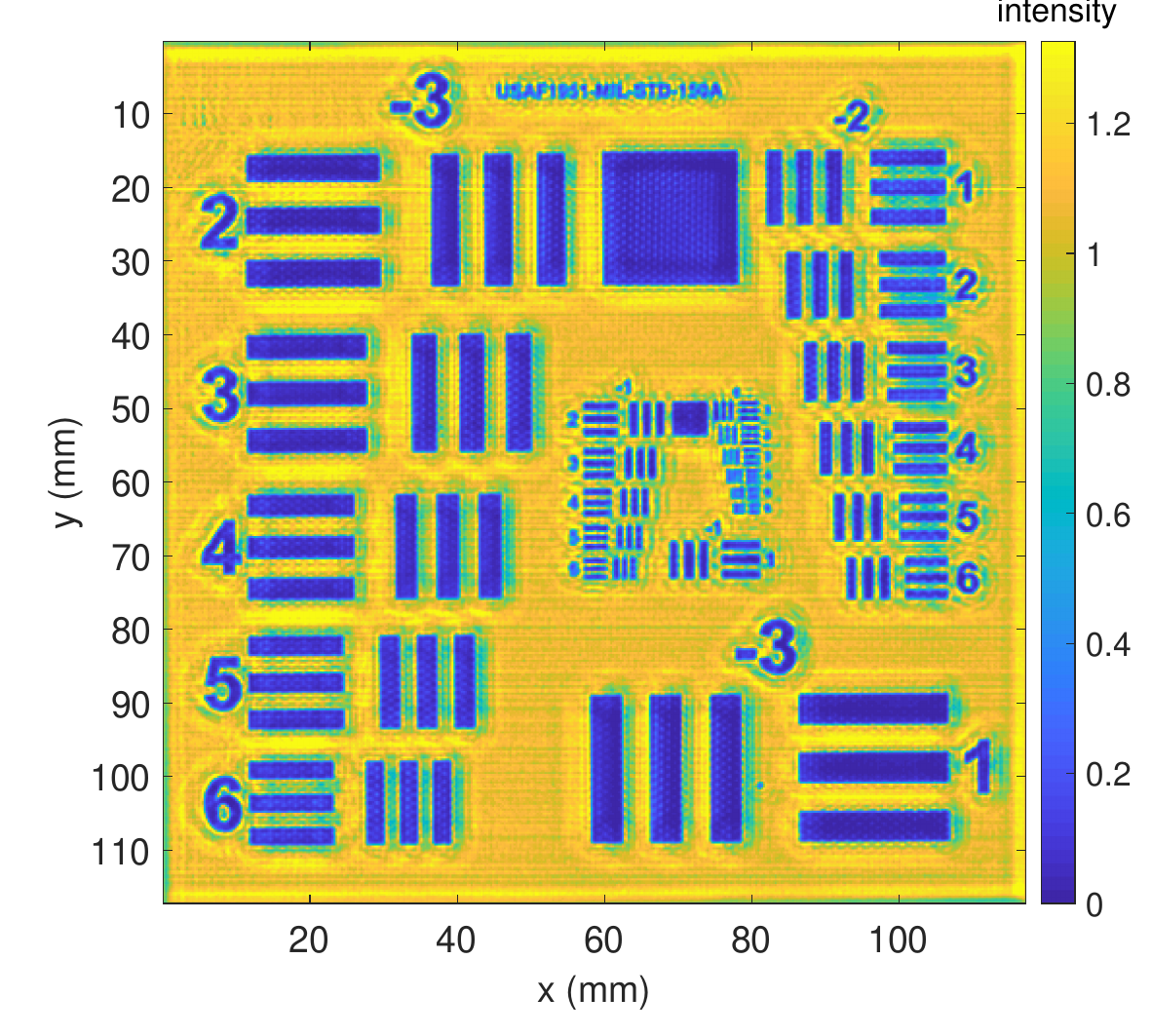}}
           
  \hfill		
  \vfill
  
  \subfloat[\label{fig:intensity_ud}\ReferXu (Xu; refer. int. $I_u$)]
           {\includegraphics[width=0.39\textwidth]{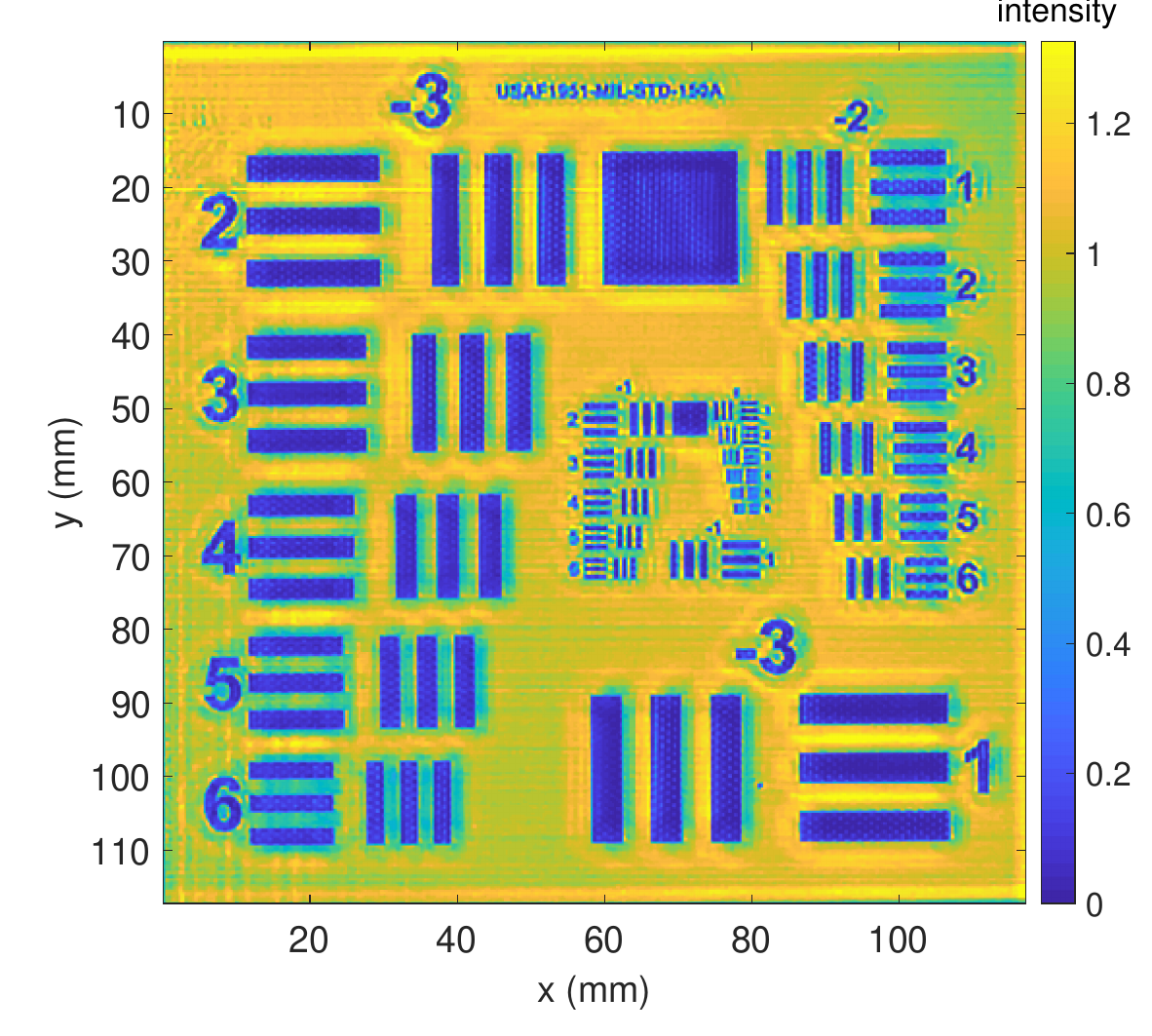}}
           \hspace{1cm}
  \subfloat[\label{fig:intensity_vd}\ReconstXu (Xu; reconst. int. $I_v$) \textbf{(proposed method)}]
           {\includegraphics[width=0.39\textwidth]{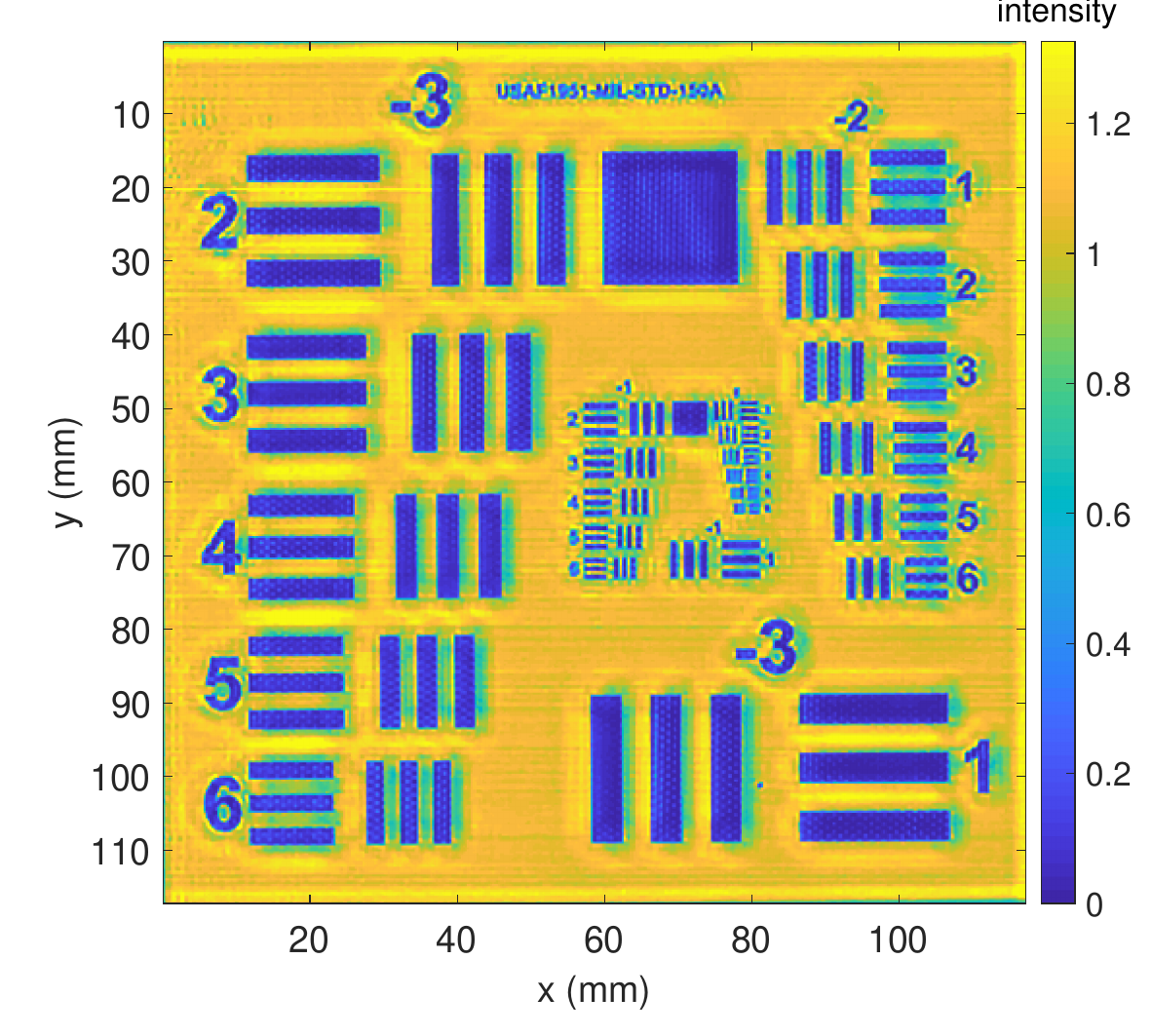}}
  
  \hfill
  \caption{Comparison between the three deconvolution methods 
  (\protect\subref{fig:intensity_u_lg}\protect\subref{fig:intensity_v_lg} Lucy-Richardson (LR)~\cite{lucy1974iterative, richardson1972bayesian} with gaussian kernel, top row ; 
  \protect\subref{fig:intensity_u_lx}\protect\subref{fig:intensity_v_lx} Lucy-Richardson with Xu's kernel, middle row ; 
  \protect\subref{fig:intensity_ud}\protect\subref{fig:intensity_vd} Xu\etal (Xu)\cite{xu2010two,xu2013unnatural}, bottom row) applied to the reference intensity image $I_u$ (left column) and the reconstructed intensity image $I_v$ (right collumn). 
  \protect\subref{fig:intensity_vd} depicts the proposed deconvolution method on our reconstructed intensity image using sparse kernel, and the PCB region for Sec.~\ref{s:evaluation.lateral.cross} is indicated.}
  \label{fig:intensity_comparison2}
\end{figure*}

\begin{table*}[!t]
\renewcommand{\arraystretch}{1.5}
\centering
\caption{Lateral resolution enhancement based on the 3~db criterion using the USAF test target}
\label{tab:lateral_resolution}
\begin{tabular}{ll|r|r|r|r}
\toprule
Method & 
Figure & 
\begin{tabular}[c]{@{}r@{}}Horizontal\\ resolution\\ ($\mu m$)\end{tabular} & 
\begin{tabular}[c]{@{}r@{}}Horizontal\\ improve-\\ ment \\ \end{tabular} & 
\begin{tabular}[c]{@{}r@{}}Vertical\\ resolution\\ ($\mu m$)\end{tabular} & 
\begin{tabular}[c]{@{}r@{}}Vertical\\ improve-\\ ment \\ \end{tabular} \\
\hline
\ReconstNoIE ($I_{v}$) & Fig.~\ref{fig:intensity_v} & 794.3 & -- & 762.0 & -- \\
\ReconstLRG  & Fig.~\ref{fig:intensity_v_lg} & 419.4 & 1.89 & 393.3 & 1.94 \\
\ReconstLRXu  & Fig.~\ref{fig:intensity_v_lx} & 402.3 & 1.97 & 368.6 & 2.07 \\
\ReconstXu ($I_{d}$)$^1$ & Fig.~\ref{fig:intensity_vd} & 346.2 & 2.29 & 359.6 & 2.12 \\
\bottomrule
\multicolumn{6}{l}{\footnotesize{$^1$ \ReconstXu ~is the proposed method, $I_{d}$}} \\
\end{tabular}
\end{table*}

In Fig.~\ref{fig:intensity_comparison2}, we show different output intensity images with respect to the input intensity image, the deconvolution method and the blur kernel. The left column shows the deconvolution results based on the reference intensity image $I_u$, whereas the right column depicts the results using the reconstructed intensity image $I_v$. For deconvolution, we apply Xu\etal's method~\cite{xu2010two,xu2013unnatural} (\ReferXu, \ReconstXu, top row) and Lucy-Richardson's original method using a gaussian kernel based on the theoretical numeric aperture (Fig.~\ref{fig:kernel_gauss}) (\ReferLRG, \ReconstLRG, center row), because the Lucy-Richardson's deconvolution method with a gaussian kernel based on the system numerical aperture is a commonly applied THz deblurring method~\cite{xu2014high}~\cite{li2008super}~\cite{ding2010high}. We, furthermore, extract the sparse kernel estimated by Xu\etal's method (Fig.~\ref{fig:kernel_sparse}) and plug it into the Lucy-Richardson approach (\ReferLRXu, \ReconstLRXu, bottom row). Note, that \ReconstXu~(Fig.~\ref{fig:intensity_vd}), is  equal to $I_d$, the method proposed in this paper.\par

In Fig.~\ref{fig:kernel_gauss} and Fig.~\ref{fig:kernel_sparse}, the gaussian kernel and Xu's sparse kernel computed by blind deconvolution are shown respectively. 
Obviously, the kernel estimated by Xu's method models further effects related to the imaging system and/or the observed target that are not covered by the gaussian kernel; we will investigate these influences in our future work. 
In Fig.~\ref{fig:kernel_cross}, we can observe that Xu's sparse kernel is not centered in the y-axis. 
This is because the blind deconvolution does not assume any symmetric and centralized property during kernel optimization, 
but instead the kernel weights are optimized as fully independent parameters throughout the process.\par

Comparing the results visually, we clearly see that \ReferLRG, \ReconstLRG~(Lucy-Richardson with gaussian kernel), 
which has previously been used for THz resolution enhancement (see Sec.~\ref{s:prior}), 
yields inferior results in terms of sharpness and local contrast. 
Xu\etal's method that estimates the blur kernel from the given intensity image yields much sharper images with improved local contrast (\ReferXu, \ReconstXu). Using Xu\etal's kernel instead of the standard gaussian kernel clearly improves the results obtained by Lucy-Richardson (\ReferLRXu, \ReconstLRXu). On the downside, Xu\etal's and Lucy-Richardson's method with Xu's kernel increase noise and overshooting effects. Xu\etal' result is, however, less affected by these artefacts. Apparently, all three methods benefit from the enhanced intensity image using our reconstruction method.\par

\begin{figure*}[!t]
  \centering
  	\hfill
  	\subfloat[\label{fig:resolution_horizontal}]{\includegraphics[width=0.49\textwidth]{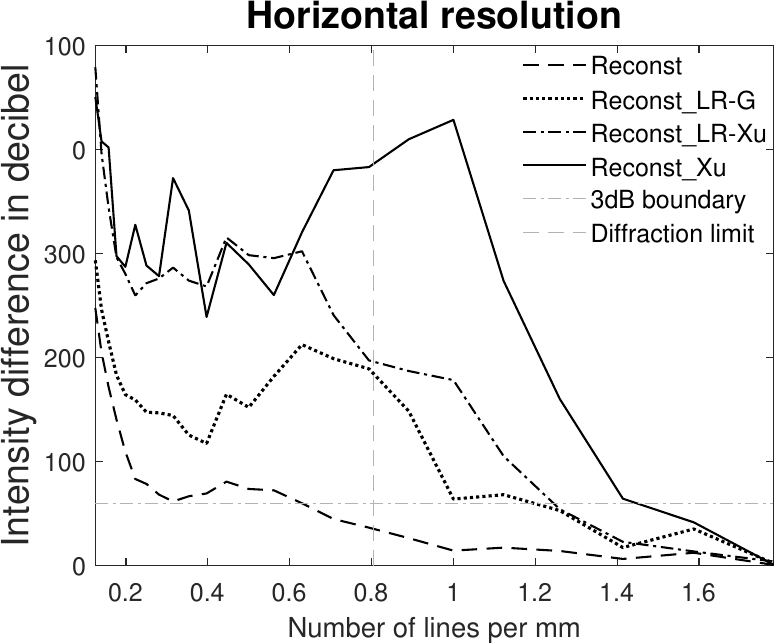}}
	\hfill
	\subfloat[\label{fig:mtf_horizontal}]{\includegraphics[width=0.49\textwidth]{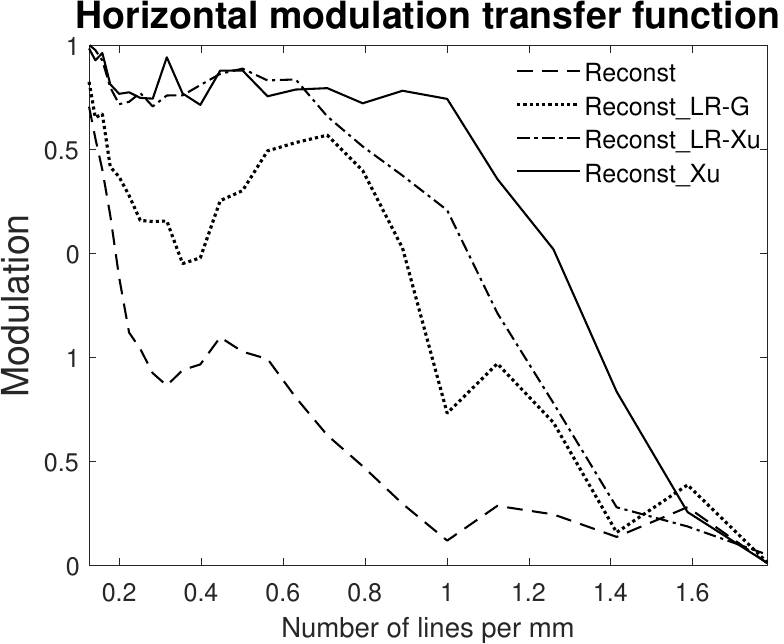}}
	\hfill
	\vfill
	\hfill
  	\subfloat[\label{fig:resolution_vertical}]{\includegraphics[width=0.49\textwidth]{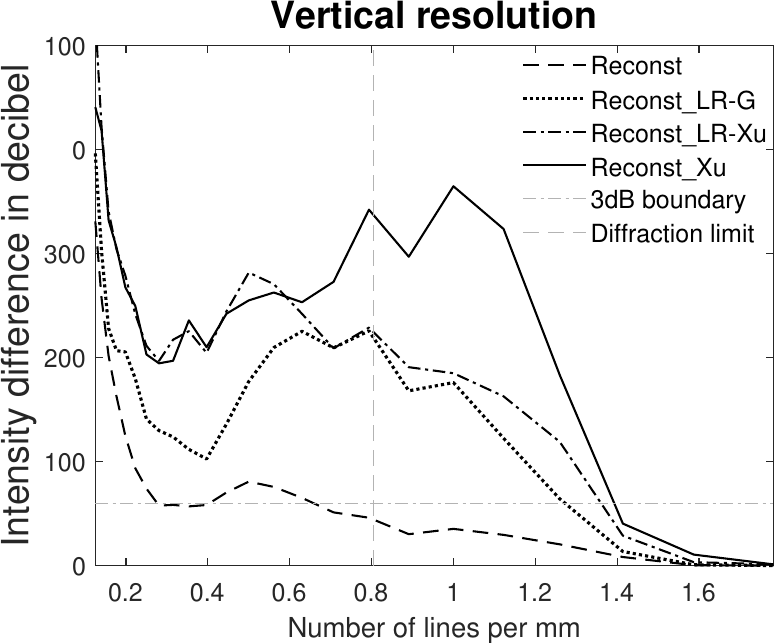}}
	\hfill
	\subfloat[\label{fig:mtf_vertical}]{\includegraphics[width=0.49\textwidth]{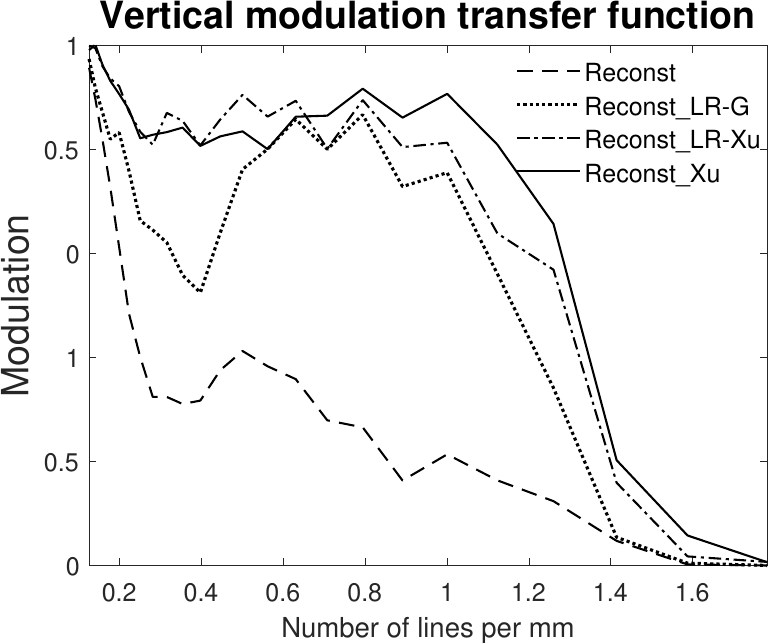}}
	\hfill
  \caption{Comparison of intensity difference and modulation transfer function (MTF) by methods to dimensions 
  	\protect\subref{fig:resolution_horizontal} horizontal intensity difference
  	\protect\subref{fig:mtf_horizontal} horizontal MTF
	\protect\subref{fig:resolution_vertical} vertical intensity difference
	\protect\subref{fig:mtf_vertical} vertical MTF}
  \label{fig:intensity_difference}
\end{figure*}

Moreover, Fig.~\ref{fig:intensity_contrast} depicts the resolution enhancement using  a cross section of group 0 element 4 of the USAF test target (see Fig.~\ref{fig:usaf_photo}), representing a line distance of $353.6\mu m$.
The cross section intensities before deconvolution (\ReconstNoIE) and after deconvolution (\ReconstLRG, \ReconstXu, \ReconstLRXu) are shown respectively 
(refer to Fig.~\ref{fig:intensity_v}, \ref{fig:intensity_v_lg}, ~\ref{fig:intensity_vd} and ~\ref{fig:intensity_v_lx}).
Only the proposed method \ReconstXu~can resolve this resolution according to the $3~dB$ criterion.
One should note that there are different definitions for resolution as discussed in~\cite{forshaw1983spatial}.
However, the center position of \ReconstLRXu, \ReconstXu~are both shifted, 
because the sparse kernel Fig.~\ref{fig:kernel_sparse} is not centered at the middle.
Although blind deconvolution does not retain the absolute intensity position, relative intensity positions are constant because we assume a spatial invariant blur kernel.\par

In order to quantify a lateral resolution, we evaluate the contrast at each of the horizontal and vertical resolution patterns of the \textbf{MetalPCB} dataset (USAF target). 
In case of vertical stripes, we determine the minimal and maximal intensity values for each row crossing the pattern's edges given its known geometric structure. 
After removing 10\% of the cross sections in the boundary area, the mean value is computed as intensity difference (in $dB$). 
Analogously, we compute the intensity difference for the horizontal stripes. 
Fig.~\ref{fig:intensity_difference} shows the intensity difference by the number of lines per mm and the modulation transfer function (MTF)~\cite{boreman2001modulation,smith1966modern} for all methods for vertical and horizontal resolutions. 
Commonly, to avoid simple enhancement by linear multiplication, a logarithmic measurement over $3~dB$ intensity difference is considered as resolution boundary. 
Tab.~\ref{tab:lateral_resolution} depicts the highest resolution at which each method delivers $\geq 3~dB$. 
We notice, that there is a resolution improvement from the non-deconvoluted image $I_v$ (\ReconstNoIE), 
via the original Lucy-Richardson method (\ReconstLRG) and the one using Xu's kernel (\ReconstLRXu), 
finally, to the proposed method (\ReconstXu). 
Taking the $3~dB$ limit as boundary, we find horizontal improvement factors (in terms of resolution) of $2.29$ for the proposed method \ReconstXu~and of $1.89$ and $1.97$ for the Lucy-Richardson \ReconstLRG~and the improved Lucy-Richardson \ReconstLRXu, respectively. 
For the vertical resolution, the proposed method has a vertical improvement factor of $2.12$, while the Lucy-Richardson methods and the improved Lucy-Richardson method have vertical improvement factor $2.07$, respectively. With respect to the modulation transfer function, Fig.~\ref{fig:mtf_horizontal} and Fig.~\ref{fig:mtf_vertical} show a significantly higher contrast for the proposed method (\ReconstXu) compared to both, the Lucy-Richardson (\ReconstLRG) and improved Lucy-Richardson (\ReconstLRXu) method.\par

\subsubsection{Embedded Structures}
\label{s:evaluation.lateral.cross}

\begin{figure}[!t]
	\centering
	\subfloat[\label{fig:intensity_u_pcb}]{\includegraphics[width=0.5\textwidth]{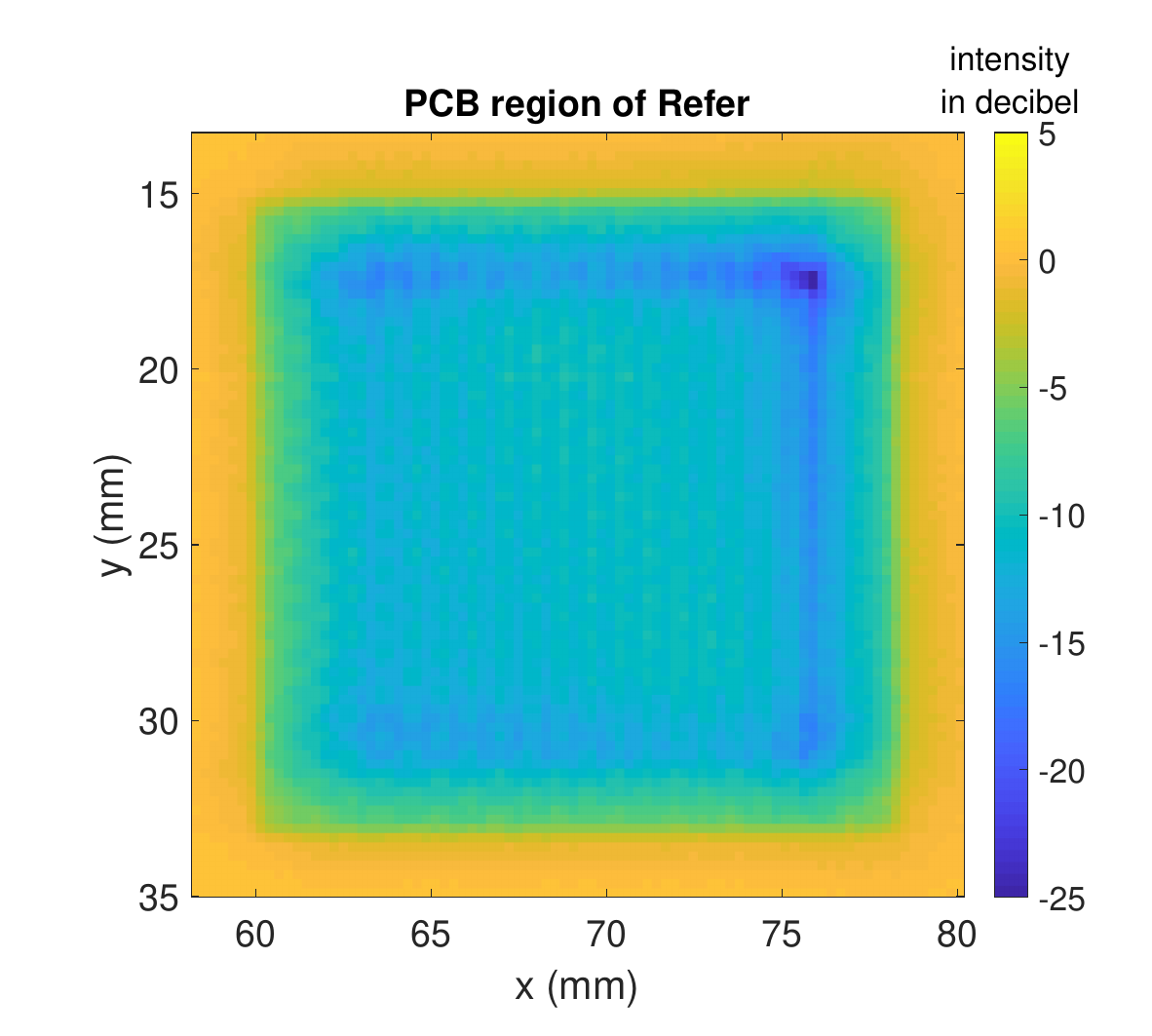}}
	\hfill
	\subfloat[\label{fig:intensity_d_pcb}]{\includegraphics[width=0.5\textwidth]{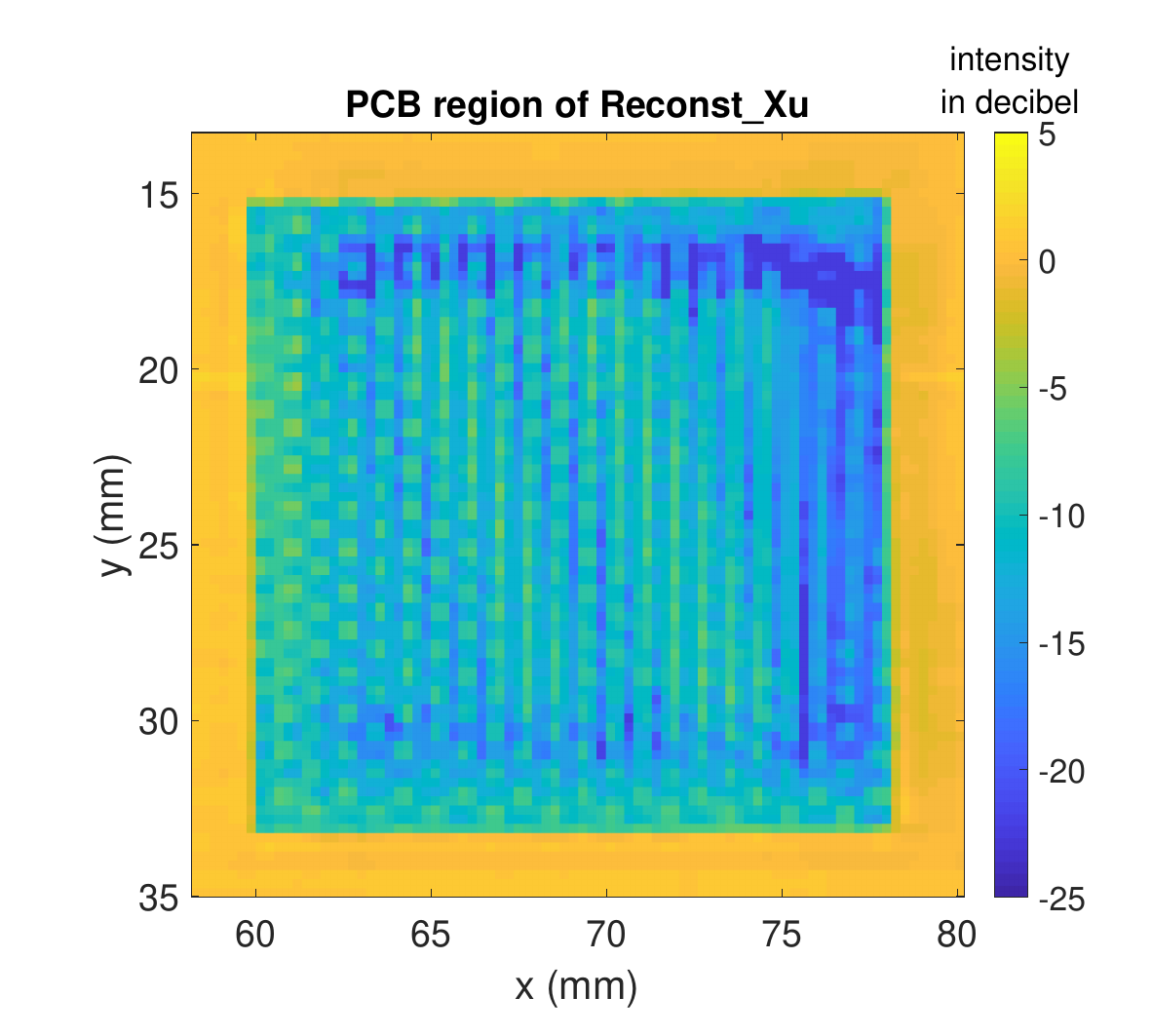}}
	\hfill
	\caption{Comparison between
	\protect\subref{fig:intensity_u_pcb} PCB region of reference intensity $I_u$ (in decibel)
	\protect\subref{fig:intensity_d_pcb} PCB region of the proposed deconvoluted intensity $I_d$ (in decibel).}
	\label{fig:intensity_comparison3}
\end{figure}

\begin{figure}[!t]
	\centering
	\includegraphics[width=0.9\textwidth]{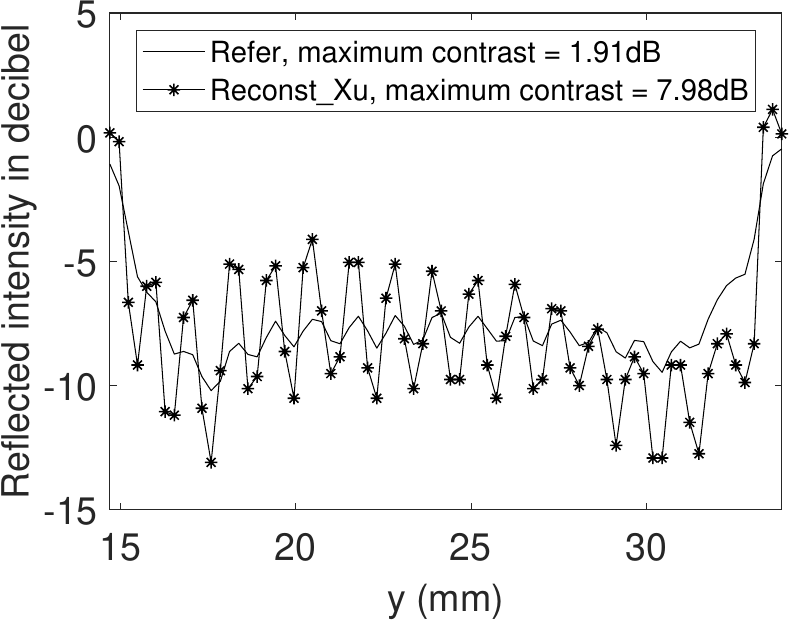}
  	\caption{Cross section in PCB region}
  	\label{fig:intensity_comparison4}
\end{figure}

In this part, we analyzed the PCB region on \textbf{MetalPCB} dataset
more closely in order to investigate the effect of our enhanced
lateral resolution on the embedded silk structures.\par

In Fig.~\ref{fig:intensity_comparison3}, the PCB region of the reference intensity $I_u$ \ReferNoIE~(see Fig.~\ref{fig:intensity_u}) and the proposed deconvoluted intensity $I_d$ \ReconstXu~(see Fig.~\ref{fig:intensity_vd}) are shown. 
We can observe that a periodic intensity pattern is visible by the enhanced lateral resolution that is caused by woven silk material embedded in the PCB region. The silk fibers  introduce a second energy reflection to the imaging system. 
In Fig.~\ref{fig:intensity_comparison4}, the cross section intensity of the PCB regions in decibel scale are shown. 
Extracting the maximum contrast within the PCB region we find an  enhancement from $1.91dB$ to $7.98dB$. Even though the proposed method nicely emphasizes the periodic silk structure underneath the polymer surface material, we cannot extract the depth of this embedded structure with our approach, as the handling of multi-reflection effects is beyond the scope of this paper. Multi-reflection effects are part of the future work.\par


\section{Conclusion}
\label{s:conclusion}

In this paper, we propose a THz computational image enhancement method to enhance the lateral resolution and depth accuracy beyond the diffraction limit. 
The method is based on a complex curve fitting on THz 3D image azimuth axis, and a blind deconvolution method on the lateral domain. 
The experiment results show that our curve fitting method enhances the depth accuracy to $91\mu m$. 
Because of this enhanced depth accuracy, the experiment also shows that the proposed reconstruction method achieves improved intensities on homogeneous, non-planar material regions without introducing additional distortions or noise. \par

Based on the reconstructed intensities, we apply several lateral blind deconvolution methods. 
Evidently, all three examined THz deconvolution methods benefit from the enhanced intensity image using our method.
In comparison to the classical Lucy-Richardson deconvolution algorithm, the experiments show that the
proposed blind deconvolution method achieves the best horizontal resolution
enhancement from $794.3\mu m$ to $346.2\mu m$, yielding an improvement
factor of $2.29$. In terms of intensity contrast, the proposed method
clearly outperforms earlier approaches.  Moreover, the experiments
show that the proposed method is able to emphasize fine
silk texture embedded within a polymer material.\par


\section*{Acknowledgement}
This research was funded by the German Research Foundation (DFG) as part of the research training group GRK 1564 \textit{Imaging New Modalities}.

This is a pre-print of an article published in Journal of Infrared, Millimeter, and Terahertz Waves. The final authenticated version is available online at: \href{https://doi.org/10.1007/s10762-019-00609-w}{https://doi.org/10.1007/s10762-019-00609-w}


\bibliographystyle{unsrt}

\end{document}